\documentclass[]{interact}

\usepackage{epstopdf}
\usepackage[caption=false]{subfig}
\usepackage{url}
\usepackage{xcolor}
\usepackage{textcomp}
\usepackage{stfloats}
\usepackage{url}
\usepackage{verbatim}
\usepackage{graphicx}
\usepackage{cite}
\usepackage{lipsum}
\usepackage{graphicx}
\usepackage{booktabs}
\usepackage{makecell}
\usepackage{color}
\usepackage{bookmark}
\usepackage{float}
\usepackage[longnamesfirst]{natbib}
\bibpunct[, ]{[}{]}{;}{n}{,}{,}
\renewcommand\bibfont{\fontsize{10}{12}\selectfont}

\theoremstyle{plain}

\theoremstyle{definition}

\theoremstyle{remark}

\begin{document}


\title{Physics-Informed Neural Network Modeling of Vehicle Collision Dynamics in Precision Immobilization Technique Maneuvers}

\author{
\name{Yangye Jiang\textsuperscript{a}, Jiachen Wang\textsuperscript{a} and Daofei Li\textsuperscript{a*}\thanks{CONTACT Daofei Li Email: dfli@zju.edu.cn}}
\affil{\textsuperscript{a}Human Mobility Automation Group, Institute of Power Machinery and Vehicular Engineering, College of Energy Engineering, Zhejiang University, Hangzhou, China}
}

\maketitle

\begin{abstract}

Accurate prediction of vehicle collision dynamics is crucial for advanced safety systems and post-impact control applications, yet existing methods face inherent trade-offs among computational efficiency, prediction accuracy, and data requirements. This paper proposes a dual Physics-Informed Neural Network framework addressing these challenges through two complementary networks. The first network integrates Gaussian Mixture Models with PINN architecture to learn impact force distributions from finite element analysis data while enforcing momentum conservation and energy consistency constraints. The second network employs an adaptive PINN with dynamic constraint weighting to predict post-collision vehicle dynamics, featuring an adaptive physics guard layer that prevents unrealistic predictions whil e preserving data-driven learning capabilities. The framework incorporates uncertainty quantification through time-varying parameters and enables rapid adaptation via fine-tuning strategies. Validation demonstrates significant improvements: the impact force model achieves relative errors below 15.0\% for force prediction on finite element analysis (FEA) datasets, while the vehicle dynamics model reduces average trajectory prediction error by 63.6\% compared to traditional four-degree-of-freedom models in scaled vehicle experiments. The integrated system maintains millisecond-level computational efficiency suitable for real-time applications while providing probabilistic confidence bounds essential for safety-critical control. Comprehensive validation through FEA simulation, dynamic modeling, and scaled vehicle experiments confirms the framework's effectiveness for Precision Immobilization Technique scenarios and general collision dynamics prediction.
\end{abstract}

\begin{keywords}
Precision immobilization technique; Post impact control; Vehicle collision dynamics; Physics-informed neural networks; Gaussian mixture model; Uncertainty quantification
\end{keywords}

\section{Introduction}
\subsection{Background and Motivation}\label{Background}

Road traffic accidents are still a significant threat to public safety. Vehicle collisions in traffic accidents often go through three dynamic stages: pre-collision, during-collision, and post-collision \cite{Parseh2023Motion}. 
Although its time duration with direct vehicle collision is very short, the second stage of `during colision' is crucial in the entire vehicle behavior progression after accidents. 
Therefore, accurately predicting the vehicle motion dynamics in collision is not only of great significance for accident reconstruction analysis and vehicle safety system design, but also a key technological foundation for developing next-generation active safety technologies, e.g. post-impact control strategies \cite{Elkady2017Collision,Yang2025Post-impact,Yang2013Vehicle}.
Vehicle rear-quarter collisions, where impacts occur at the vehicle's rear corner area as in Precision Immobilization Technique (PIT) maneuvers, represent one of the most complex and dynamically challenging impact scenarios in traffic safety research \cite{Zhou2008Vehicle,Zhou2008Collision,Mascarenas2017Autonomous}. 
The research of PIT technique and post-impact control systems have created new demands for accurate, real-time collision prediction capabilities that traditional methods struggle to meet. 
With advancing research in vehicle active safety control, there is a growing urgent need for a modeling approach that balances accuracy and computational efficiency to support model-based control systems, enhance post-impact reliability, and promote practical applications.

\subsection{Literature Review}\label{class}
\subsubsection{Impact Force Modeling Methods}

To better model vehicle post-impact dynamics, it is necessary to understand the vehicle behavior changes during the collision process, which are intuitively reflected in the impact force curve. For lite collisions such as PIT maneuvers, the force curve can be estimated from accelerometer data \cite{Huibers2001current}. 
Huibers et al. \cite{Huibers2001current} found that for collisions with the same barrier type, the trend of force curve $F(x)$ remains consistent across different vehicle impacts. However, the magnitude of impact force varies depending on specific operational condition and the structural characteristic of the vehicle involved.

Various modeling methods for impact force have been developed. Lumped parameter models based on spring-damper systems \cite{Jonsen2009Identification, Ofochebe2015Performance} and pulse models \cite{Iraeus2015Pulse, Kim2014Optimal} represent common approaches. 
Pulse approximation methods, including haversine, half-sine, and triangular waveforms, have been widely adopted in the vehicle safety domain \cite{Huang2002, Wei2017Data-based}. 
These simplified mathematical formulations provide computationally efficient representations of impact forces, enabling rapid analysis and preliminary design evaluations while maintaining acceptable accuracy for certain crash scenarios. 
For more detailed analysis, multibody dynamics models in the discrete time domain \cite{Sousa2008Development} and finite element models \cite{Chen2021deepa, Yildiz2012Multi-objective} can be employed. 
These approaches enable comprehensive investigation of vehicle behavior in terms of deformation, displacements, velocity, and accelerations throughout the entire impact duration. 
However, the substantial computational demand limits their application to analyzing vehicle dynamics and modeling transient collision behaviors, rendering them unsuitable for direct use in developing vehicle active safety algorithms. 
Nevertheless, the data generated by these methods can still, to some extent, substitute for real-world collision data that are prohibitively expensive to obtain.

Chen et al. \cite{Chen2021deepa} developed a deep neural network-based approach for solving inverse problems in vehicle collisions using finite element data, achieving prediction accuracy with mean errors below 3\%. Yildiz and Kiran \cite{Yildiz2012Multi-objective} applied the finite element method to particle swarm-based optimization for vehicle crash safety.

\subsubsection{Post-impact Vehicle Dynamics Modeling}
 
Vehicle post impact motion prediction faces multiple challenges. Firstly, the collision process involves complex nonlinear dynamic phenomena, including the coupled effects of various factors such as structural deformation, changes in tire mechanical properties, and suspension system responses \cite{Vangi2020Vehicle}. 
Secondly, post-impact vehicle motion often exhibits large sideslip angles and yaw rates that far exceed the dynamic response range under normal driving conditions \cite{Wang2022Post-Impact}. 
Furthermore, in special collision scenarios such as PIT maneuvers, the vehicle dynamic response becomes even more complex, making it difficult for traditional vehicle dynamics models to accurately describe its motion patterns. 
Research by the Michigan State Police has revealed that performing PIT maneuvers on vehicles equipped with Electronic Stability Control (ESC) systems can yield unpredictable outcomes at both low and high speeds, whereas vehicles without ESC demonstrate significantly more consistent and predictable responses \cite{OJP2024}.

Existing collision modeling methods face inherent limitations. Momentum conservation approaches, such as the Kudlich-Slibar method, offer high computational efficiency but assume instantaneous collision and neglect critical physical phenomena during impact, such as tire forces \cite{Brach1977Impact, Ishikawa1993Impact, Kolk2016Evaluation,Emori1970, Brach1983analysis}. 
To address these issues, Zhou et al. \cite{Zhou2008Vehicle, Zhou2008Collision} developed a multi-stage vehicle dynamics model for PIT, quantitatively analyzing the influence of operational parameters on impact outcomes. 
Elmarakbi et al. \cite{Elmarakbi2013Numerical} simulated vehicle oblique frontal collisions using a 6-degree-of-freedom mathematical model, while Gidlewski et al. \cite{Gidlewski2019process} modeled the dynamics and energy balance of vehicle frontal-to-side collisions. 
However, these studies rely on oversimplified linear tire slip angle models, which fail to accurately capture actual vehicle responses. 
Kim et al. \cite{Kim2014Optimal} introduced an effective factor to approximate the nonlinear region of tire behavior, enabling a 4DOF model to describe vehicle responses under collision forces and validate against CarSim simulations. 
Nevertheless, this method still requires extensive precise model parameters and calibration based on tire force data, which is impractical in emergency situations characterized by high uncertainty.

\subsubsection{Vehicle Dynamic Modeling Based on Data-driven Methods}


In recent years, with the rapid development of artificial intelligence technology, especially the successful application of deep learning in various fields, machine learning methods have demonstrated great potential in vehicle dynamics modeling \cite{DaLio2020Modelling, Chrosniak2023Deep}. 
Spielberg et al. \cite{Spielberg2019Neural} developed a high-performance autonomous driving vehicle model based on neural networks, using a feedforward-feedback control architecture and historical state information to handle vehicle dynamics under different road friction conditions. 
Williams et al. \cite{WilliamsInformation} combined the model predictive path integral (MPPI) algorithm and a neural network-learned dynamics model to achieve real-time control of complex nonlinear systems. 
Pan et al. \cite{Pan2021Data-driven} integrated complex multibody dynamics simulation with deep learning to achieve real-time vehicle dynamics prediction. 
These methods utilize large amounts of data for neural network-based vehicle dynamics modeling, particularly the nonlinear tire mechanics in vehicles \cite{Spielberg2019Neural,Kim2022Physics}. 
However, these methods rely on extensive training data, and lack physical model transparency and intuitiveness, while the trained models can only be used for specific vehicle types and operating conditions.
Although purely data-driven neural network models excel at learning complex nonlinear patterns, they often fail to guarantee consistency with physical constraints, potentially generating physically unrealizable state prediction for the vehicle.


To overcome the limitations of purely data-driven methods, Physics-Informed Neural Networks (PINN) have emerged as an effective solution. PINN represents a class of universal function approximators that embed physical laws, formulated as partial differential equations, into the neural network learning process \cite{Karniadakis2021Physics-informed, Raissi2019Physics-informed}. 
Unlike methods that rely on data generated from physical models \cite{Spielberg2019Neural}, PINN incorporates physical prior knowledge as regularization terms to constrain the solution space of acceptable functions, thereby enhancing the generalization capability of function approximation even under data-scarce conditions. 
This advantage is particularly evident for simple vehicle models \cite{Mo2021physics-informed, Shi2023Physics-informed,Cheng2025hybrid,Geng2023physics-informed}. 
Such approach is especially suitable for engineering scenarios where data availability is limited. For example, Tan et al. \cite{Tan2025Modeling} utilize PINN to model the dynamics of experimental vehicles. 
To address the challenge of neural network convergence with limited data, Fang et al. \cite{Fang2024Fine} proposed a Fine-Tuning Hybrid Dynamics (FTHD) method, which fine-tunes a pre-trained Deep Dynamics Model (DDM) using a small training dataset.

Similarly, the PINN-based approach proposed by Long et al. \cite{Long2024Physics-informed} enhances conventional neural networks by incorporating physical prior knowledge such as vehicle kinematics and boundary conditions. These methods demonstrate excellent extrapolation capability across dynamic scenarios, enabling the reconstruction of high-speed trajectories using only low-speed training data. However, their limitations primarily lie in excessive reliance on physical models, restricting their applicability to relatively simple scenarios and making them inadequate for characterizing uncertainties in complex dynamic environments.


\subsubsection{Research Gaps}
Despite significant advances in both collision modeling and physics-informed learning, several critical gaps remain:

\begin{enumerate}
  \item Existing neural network architectures struggle to effectively handle the strong discontinuities of contact and force during collisions.
  \item Collisions involve multiple time scales, from millisecond impact duration to seconds-long post-collision motion. Most existing studies can only simplify the impact process based on experience \cite{Parseh2023Motion,Zhou2008Collision}.
  \item Traditional neural networks require extensive training data, which is costly and dangerous to obtain for collision scenarios.
  \item To address security issues, existing models struggle to balance computational complexity and accuracy.
\end{enumerate}

Based on the current research status and existing challenges, this paper proposes two PINN-based neural networks. The first network integrates finite element analysis (FEA) data with baseline physical models to model millisecond-level impact forces, while the second network incorporates fine-tuning techniques to achieve accurate modeling of post-collision vehicle attitude changes. Both models integrate gaussian mixture models (GMM) to handle uncertain state distributions. The first model is validated using a self-constructed FEA database, demonstrating its capability for accurate impact force modeling. The second model is validated through a combined FEA-CarSim simulation environment and scaled vehicle experiments.

\subsection{Contributions}

This paper addresses these gaps through the following contributions:

\begin{enumerate}
  \item We propose a dual physics-informed neural network framework where the first network models time-varying collision force distributions and the second predicts post-collision vehicle dynamics, achieving complete collision process modeling across temporal scales.
  \item We design physics guard layer that balances physical consistency with data learning through dynamic constraint weight adjustment and soft boundary constraints, effectively addressing the issue of physical constraints suppressing neural network learning.
  \item We establish a multi-fidelity validation framework from FEA to CarSim simulation and scaled vehicle experiments, reducing trajectory prediction error by 63.6\% compared to traditional 4DOF models while maintaining millisecond-level computational efficiency.
\end{enumerate}

\section{Baseline Vehicle Collision Dynamics Modeling}
\subsection{Vehicle Dynamics Model Considering Impact Forces}
\begin{figure}[h]
  \centering
  \includegraphics[width=8cm]{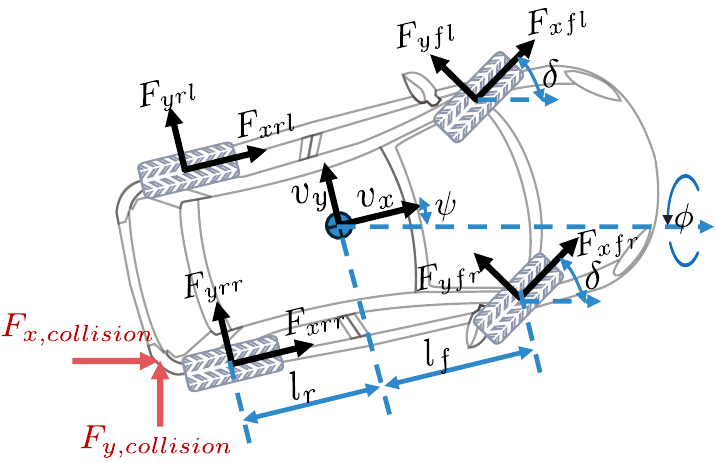}
  \caption{Schematic diagrams of the 4DOF vehicle model with impact forces applied}
  \label{fig:vehicle_model}
\end{figure}

Since PINN training requires incorporating physical constraints, a continuous baseline model must first be established to provide the necessary physical foundations for the vehicle dynamics modeling framework. As shown in Figure \ref{fig:vehicle_model}, to capture essential collision dynamics while maintaining computational tractability, we adopt a four-degree-of-freedom (4DOF) vehicle model that extends traditional planar approaches by including roll motion, which is critical for accurate representation of vehicle behavior during high-acceleration maneuvers and collision events \cite{Kim2014Optimal}. The state vector is defined as:

\begin{equation}
  \label{eq:state_vector}
  \mathbf{x} = [v_x, v_y, \psi, \dot{\psi}, \phi,\dot{\phi},  X, Y]^T \in \mathbb{R}^8
\end{equation} 
where $ v_x $ and $ v_y $ are longitudinal and lateral velocities, $ \dot{\psi} $ is yaw rate, $ \dot{\phi} $ is roll rate, $ \phi $ is roll angle, $ (X, Y) $ is global position, and $ \psi $ is heading angle.

The governing equations follow the lateral-yaw-roll model with impact force augmentation, incorporating the coupling effects between lateral, yaw, and roll dynamics:
\begin{equation}
  \label{eq:4dof}
  \begin{aligned} &m\dot{v}_x = mv_y\dot{\psi} + F_{x,\text{tire}} + F_{x,\text{collision}} \\ &m\dot{v}_y = -mv_x\dot{\psi} + F_{y,\text{tire}} + F_{y,\text{collision}} \\ &I_{zz}\ddot{\psi} = M_{z,\text{tire}} + M_{z,\text{collision}} \\ &I_{xx,s}\ddot{\phi} + I_{xz}\ddot{\psi} = m_s g h_{\text{rc}} \sin\phi - K_s\phi - B_s\dot{\phi} + M_{x,\text{collision}} 
  \end{aligned} 
\end{equation} 
where $ m $ is vehicle mass, $ I_{zz} $ and $ I_{xx,s} $ are yaw and roll moments of inertia, $ I_{xz} $ is product of inertia representing the coupling between yaw and roll motions, $ K_s $ and $ B_s $ are roll stiffness and damping coefficients, $ m_s $ is sprung mass, and $ h_{\text{rc}} $ is roll center height. 

The resultant tire forces and moments are computed as the sum of contributions from all four wheels:
\begin{equation}
  \label{eq:4dof2}
  \begin{aligned} 
  &F_{x,\text{tire}} = F_{x,fl} + F_{x,fr} + F_{x,rl} + F_{x,rr} \\ &F_{y,\text{tire}} = F_{y,fl} + F_{y,fr} + F_{y,rl} + F_{y,rr} \\ &M_{z,\text{tire}} = l_f(F_{y,fl} + F_{y,fr}) - l_r(F_{y,rl} + F_{y,rr}) + \frac{t_w}{2}(F_{x,fr} - F_{x,fl} + F_{x,rr} - F_{x,rl}) 
  \end{aligned}
\end{equation} 
where $ l_f $ and $ l_r $ are distances from the center of gravity to the front and rear axles, respectively, and $ t_w $ is track width. 

The individual tire forces are modeled using a simplified magic formula tire model \cite{Parseh2023Motion}, which provides a good balance between computational efficiency and physical accuracy. The lateral forces are expressed as:
\begin{equation}
  \label{eq:tire_model}
  F_{y,i} = -\sin\left(\tan^{-1}(C_i \alpha_i)\right) \sqrt{(\mu_s F_{zi})^2 - F_{xi}^2}, \quad \text{with } i = \{fl, fr, rl, rr\}
\end{equation} 
where $ i \in {fl, fr, rl, rr} $ are tire position (front-left, front-right, rear-left, rear-right), $\mu_s$ is tire-road friction coefficient and slip angles $\alpha_i$ incorporate the effects of vehicle kinematics and steering input:
\begin{equation}
  \label{eq:tire_model2}
  \begin{aligned} 
    &\alpha_{fl} = \delta - \arctan\left(\frac{v_y + l_f \dot{\psi}}{v_x - \frac{w}{2}\dot{\psi}}\right) \\   &\alpha_{fr} = \delta - \arctan\left(\frac{v_y + l_f \dot{\psi}}{v_x + \frac{w}{2}\dot{\psi}}\right) \\ &\alpha_{rl} = -\arctan\left(\frac{v_y - l_r \dot{\psi}}{v_x - \frac{w}{2}\dot{\psi}}\right) \\ &\alpha_{rr} = -\arctan\left(\frac{v_y - l_r \dot{\psi}}{v_x + \frac{w}{2}\dot{\psi}}\right)
  \end{aligned}
\end{equation} 
where $ \delta $ is the front wheel steering angle. While longitudinal forces can generally be mapped from driver pedal inputs, for the specific PIT scenarios examined in this paper where drivers do not apply acceleration or deceleration inputs post-collision, the longitudinal force is simplified to rolling resistance only:
\begin{equation}
  \label{eq:tire_model3}
  F_{x,i} = (f_{r0} + f_{r1} \cdot v + f_{r2} \cdot v^2) \cdot F_{z,i}
\end{equation} 
where $f_r$ is the rolling friction coefficient, $f_{r0}$ denotes the basic rolling friction coefficient, $f_{r1}$ and $f_{r2}$ are velocity-dependent coefficients, and $v$ is the vehicle's longitudinal speed. The vertical load distribution is crucial for accurate tire force prediction and accounts for both longitudinal and lateral load transfer effects caused by vehicle accelerations and body roll:
\begin{equation}
  \label{eq:tire_model4}
  \begin{aligned}
    F_{z,fl} &= \frac{m}{2} \left( \frac{g l_r}{l_f + l_r} - \frac{h_{cog} a_x}{l_f + l_r} \right) - \frac{K_f}{K_f + K_r} \left( \frac{m h_{cog} a_y}{w} + \frac{m_s g h_{rc} \sin \phi}{w} \right) \\
    F_{z,fr} &= \frac{m}{2} \left( \frac{g l_r}{l_f + l_r} - \frac{h_{cog} a_x}{l_f + l_r} \right) + \frac{K_f}{K_f + K_r} \left( \frac{m h_{cog} a_y}{w} + \frac{m_s g h_{rc} \sin \phi}{w} \right) \\
    F_{z,rl} &= \frac{m}{2} \left( \frac{g l_f}{l_f + l_r} + \frac{h_{cog} a_x}{l_f + l_r} \right) - \frac{K_r}{K_f + K_r} \left( \frac{m h_{cog} a_y}{w} + \frac{m_s g h_{rc} \sin \phi}{w} \right) \\
    F_{z,rr} &= \frac{m}{2} \left( \frac{g l_f}{l_f + l_r} + \frac{h_{cog} a_x}{l_f + l_r} \right) + \frac{K_r}{K_f + K_r} \left( \frac{m h_{cog} a_y}{w} + \frac{m_s g h_{rc} \sin \phi}{w} \right) \\
  \end{aligned}
\end{equation} 
where $h_{cog}$ is the height of the center of gravity; $a_x$ and $a_y$ are the longitudinal and lateral accelerations; $K_f$ and $K_r$ are the front and rear cornering stiffnesses; $w$ is the vehicle track width; $m_s$ is the sprung mass; $h_{rc}$ is the height of the roll center; and $\phi$ is the roll angle.

\subsection{Momentum-conservation-based Collision Model}
Vehicle collision processes are fundamentally momentum transfer phenomena, with their physical foundation rooted in the momentum conservation principle from Newton's laws of motion. In collision systems where external forces are negligible compared to impact forces, the total system momentum remains conserved before and after collision. This principle provides a theoretical foundation for deterministic impulse calculations, offering higher certainty compared to directly modeling the complex nonlinear processes of impact forces.

The momentum conservation-based collision modeling approach was first systematically developed by Brach et al. \citep{Brach1977Impact}, treating colliding vehicles as rigid bodies and determining collision impulse through analysis of momentum changes before and after impact. Compared to direct impact force modeling, this approach offers advantages: impulse calculations primarily depend on relatively deterministic physical quantities such as collision geometry parameters, vehicle masses, and relative velocities, while complex factors like material nonlinearity and structural deformation have relatively minimal impact on impulse calculations.

\begin{figure}[h]
  \centering
  \includegraphics[width=10cm]{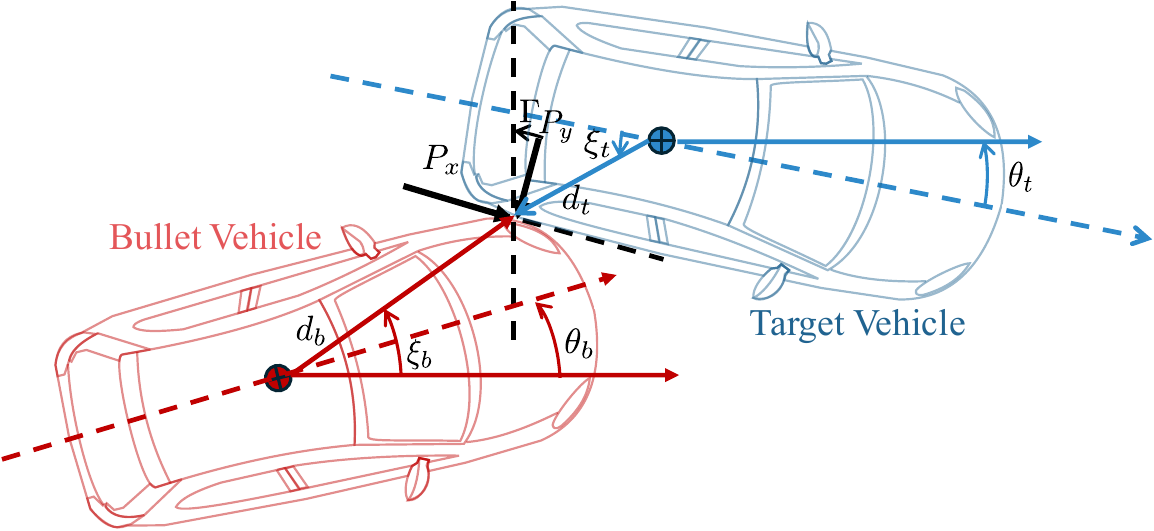}
  \caption{A view of two-vehicle collision model}
  \label{fig:collision_model}
\end{figure}

Consider a two-dimensional planar problem of two-vehicle collision. A ground-fixed coordinate system XOY is established, where the X-axis is along the tangential direction of the road. As shown in Figure \ref{fig:collision_model} Let the target vehicle be target vehicle and the striking vehicle be bullet vehicle. The pre-collision vehicle motion states are described by six parameters: ${v_{tx}, v_{ty}, \dot{\psi}_{t}, v_{bx}, v_{by}, \dot{\psi}_{b}}$, and the corresponding post-collision motion states are: ${V_{tx}, V_{ty}, \dot{\Psi}_{t}, V_{bx}, V_{by}, \dot{\Psi}_{b}}$.

According to the law of momentum conservation, the system's linear momentum is conserved in both X and Y directions as
\begin{equation}
  \label{eq:collision_model1}
  \begin{aligned} 
   &m_t \cdot (V_{tx} - v_{tx}) = -m_b \cdot (V_{bx} - v_{bx}) = P_x \\
   &m_t \cdot (V_{ty} - v_{ty}) = -m_b \cdot (V_{by} - v_{by}) = P_y
  \end{aligned}
\end{equation} 
where $P_x$ and $P_y$ are the collision impulses in the X and Y directions, respectively.
For angular momentum conservation, with each vehicle's center of mass as the moment center:

\begin{equation}
  \label{eq:collision_model2}
  \begin{aligned} 
   &I_{zzt}(\dot{\Psi}_{t} - \dot{\psi}_{t}) = P_x d_t \sin(\theta_t + \xi_t) - P_y d_t \cos(\theta_t + \xi_t) \\
   &I_{zzb}(\dot{\Psi}_{b} - \dot{\psi}_{b}) = P_x d_b \sin(\theta_b + \xi_b) - P_y d_b \cos(\theta_b + \xi_b) 
  \end{aligned}
\end{equation} 

To solve for the impulses $P_x$ and $P_y$, two additional physical constraint conditions are introduced.The restitution coefficient $e$ is defined as the negative ratio of the relative separation velocity after collision to the relative approach velocity before collision:
\begin{equation}
  \label{eq:restitution_coefficient}
  e = -\frac{V_{bn} - V_{tn}}{v_{bn} - v_{tn}}
\end{equation} 
where the subscript $n$ denotes normal component of the collision contact surface. 
The tangential friction coefficient $\mu$ is defined as the ratio of tangential impulse to normal impulse:
\begin{equation}
  \label{eq:friction_coefficient}
  \mu = \frac{P_t}{P_n}
\end{equation} 

Decomposing the impulse into the normal-tangential coordinate system yields: 
\begin{equation}
  \label{eq:friction_coefficient2}
  \mu \cdot (P_x \cos\Gamma + P_y \sin\Gamma) = P_y \cos\Gamma - P_x \sin\Gamma
\end{equation} 
By solving coupled equations \ref{eq:collision_model1}-\ref{eq:friction_coefficient2}, the collision impulses $P_x$ and $P_y$ can be determined. A key characteristic of this solution process is that the impulse calculation primarily depends on the following relatively deterministic physical parameters: geometric parameters (collision point location, vehicle dimensions, collision angle, etc.) can be accurately determined through collision geometry analysis; inertial parameters (vehicle mass, moment of inertia, etc.) possess high certainty; kinematic parameters (pre-collision velocity states) can be obtained through trajectory analysis or sensor data; and material parameters (coefficient of restitution and friction coefficient), while exhibiting some variability, have relatively limited impact and can be determined through empirical data. In contrast, directly modeling the impact force process requires considering numerous uncertain factors such as material nonlinearity, structural deformation, and contact stiffness variations, making impulse calculation inherently more deterministic. The uncertainties in practical applications will be implicitly incorporated in the GMM impact forces training discussed below.According to the impulse-momentum theorem, impulse equals the integral of force over time
\begin{equation}
  \label{eq:impulse-momentum}
  \begin{aligned} 
  &P_x = \int_0^{\Delta t} F_x(t) dt\\
  &P_y = \int_0^{\Delta t} F_y(t) dt
  \end{aligned}
\end{equation} 

Based on the determined impulses $P_x$ and $P_y$, simplified force-time models can be employed to represent the temporal distribution of impact forces. Here we used a GMM model.

\section{Impact Force Modelling Considering Uncertainties}
\subsection{FEA Environment Setting}
To establish a reliable foundation for the stochastic impact force model, this paper conducted systematic simulation research using LS-DYNA R12.0 FEA software. LS-DYNA is one of the most advanced explicit nonlinear FEA software packages internationally, with extensive applications and validation in automotive crash simulation. The software employs explicit time integration algorithms that effectively handle complex problems including large deformation, material nonlinearity, and contact collisions, making it particularly suitable for high-speed impact and collision analysis.

\begin{figure}[h]
  \centering
  \includegraphics[width=14cm]{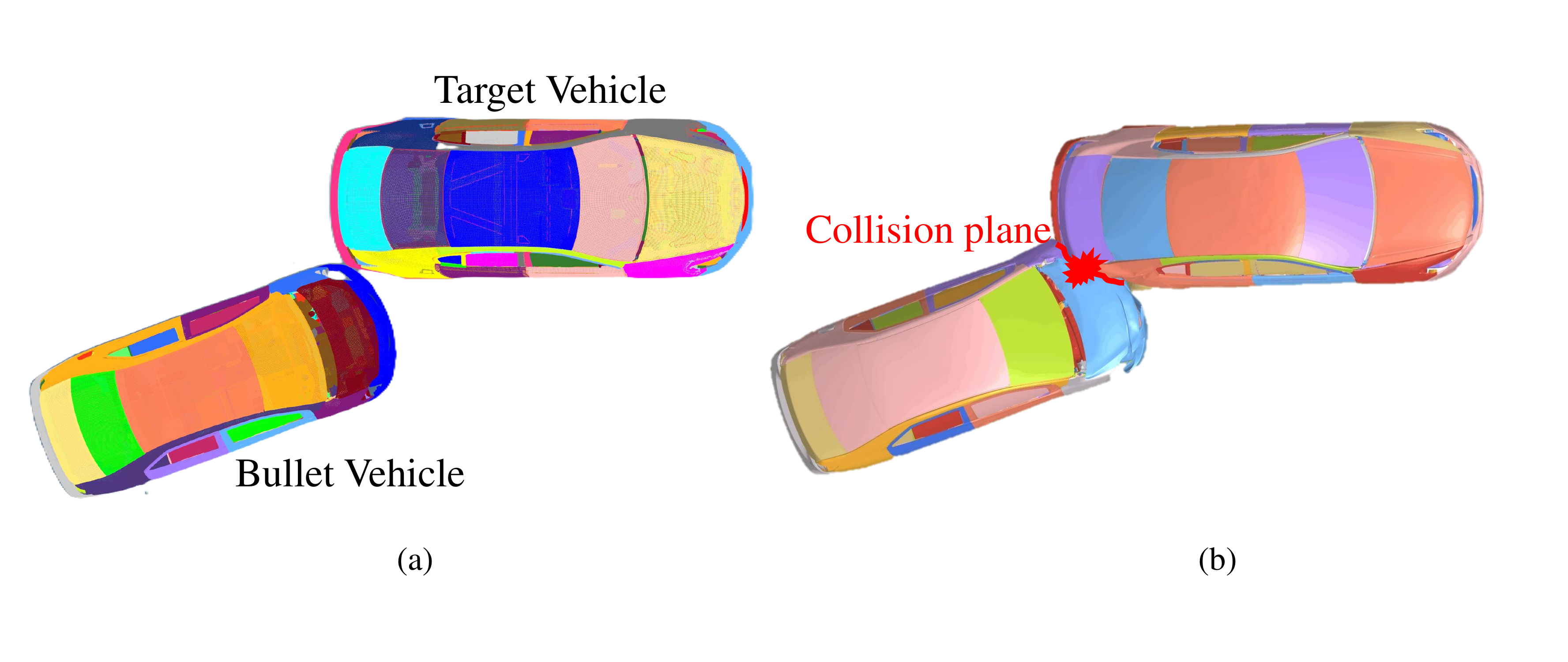}
  \caption{The PIT scenario setting in FEA simulation: (a) before and (b) after collision}
  \label{fig:lsdyna}
\end{figure}

The FEA scenarios designed in this paper are shown in the Figure \ref{fig:lsdyna},
(a) is the grid diagram and (b) is the diagram of the FEA simulation collision, where a vehicle approaches from the rear-quarter and collides with the target vehicle at the rear axle body location. The target vehicle is set to 2000kg, while the bullet vehicle considers three different mass conditions: 1500kg/2000kg/2500kg. High-fidelity vehicle models are employed, with two different vehicle models selected. Each simulation adopts rigorous numerical settings to ensure computational accuracy and stability. Time integration uses the central difference scheme with a time step of $1\times 10^{-6}$ seconds, which can capture high-frequency phenomena during collision while maintaining numerical stability. Spatial discretization employs mixed hexahedral and tetrahedral meshes, with mesh sizes in critical regions controlled at 2-5mm to accurately capture stress gradients and deformation details.

\begin{figure}[H]
  \centering
  \includegraphics[width=13cm]{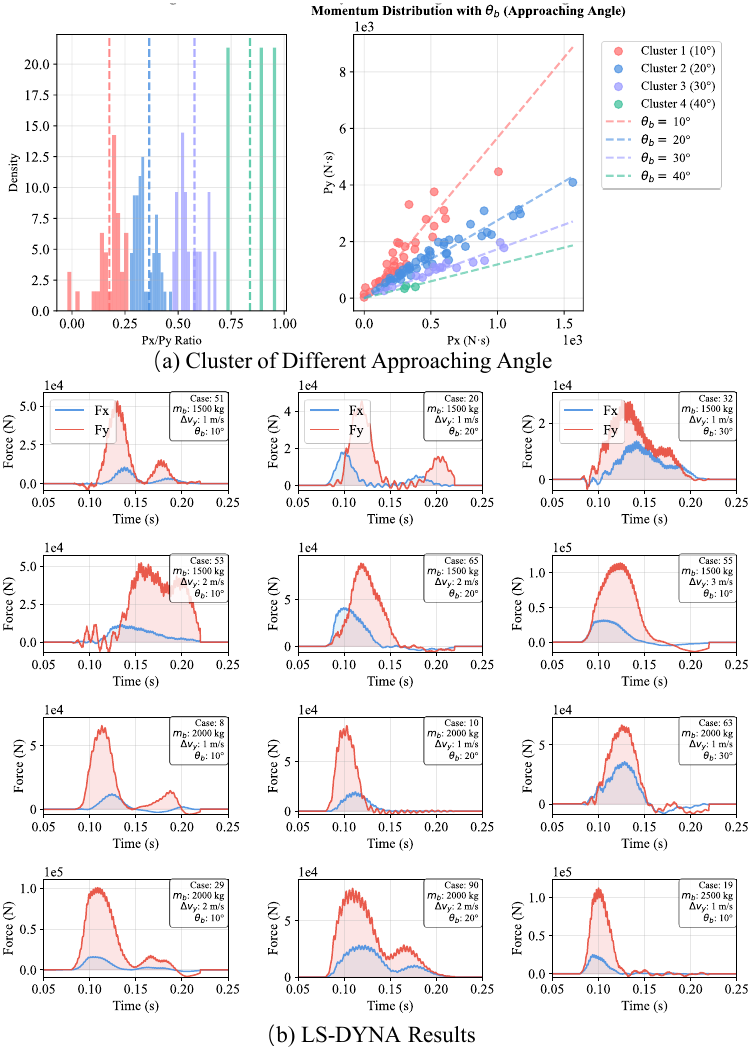}
  \caption{Twelve sample results of FEA force calculated via $a_x$ and $a_y$}
  \label{fig:lsdyna_cases}
\end{figure}

As shown in Figure \ref{fig:lsdyna_cases}, 6 cases with different parameter distributions are sampled from the 100 cases. Figure \ref{fig:lsdyna_cases} (a) presents the different parameter distributions, revealing that the values of $P_x/P_y$ under different conditions can be essentially classified by the collision angle. As shown in Figure \ref{fig:lsdyna_cases}(b), since the collision scenarios are primarily side-impact collisions, the actual $P_y$ is generally larger than $P_x$.

\subsection{PINN Based Impact force Modeling Considering Uncertainties}
Traditional approaches to impact force modeling typically rely on preset geometric profiles (e.g., triangular or rectangular pulses). While computationally efficient, such methods exhibit notable limitations in handling complex collision scenarios: (1) fixed geometric configurations fail to accommodate the inherent variations in force time-histories under differing collision conditions; (2) these methods cannot accurately predict vehicle state transitions under collision when serving as inputs for subsequent collision dynamics analyses; (3) quantification of uncertainty in prediction outcomes remains challenging.

To address these challenges, this paper proposes an innovative modeling framework that abandons traditional basis function methods, adopting a hybrid strategy integrating Physics-Informed Neural Networks (PINN) and Gaussian Mixture Models (GMMs). The core idea is to enable the neural network to autonomously learn the temporal distribution patterns of impact forces under strict physical constraints.

The complete PINN-GMM modeling framework can be expressed as

\begin{equation}
  \label{eq:pinn-gmmframework}
  \begin{aligned} 
  \mathbf{F}_{\text{collision}}(t; \boldsymbol{\theta}, \mathbf{W}) =\mathcal{NN}_{\text{PINN-GMM}}(\boldsymbol{\theta}, t; \mathbf{W})
  \end{aligned}
\end{equation} 
where $\mathbf{F}_{\text{collision}}(t) = [F_x(t), F_y(t)]^T$ denotes the impact force vector at time $t$, $\boldsymbol{\theta}$ represents the collision scenario feature vector, $\theta = [m_t,m_b,v_{ty},v_{by},v_{tx},v_{bx},\theta_b]$, $\mathbf{W}$ stands for the set of neural network weight parameters, and $\mathcal{NN}_{\text{PINN-GMM}}$ indicates the neural network mapping function integrated with physical constraints.

The key innovation of this framework resides in embedding physical laws as hard constraints into the network training process to ensure the physical rationality of prediction results, while quantifying prediction uncertainty through the GMM component. Compared with traditional methods, this framework exhibits three distinct features: (1) integration of physical laws as hard constraints; (2) probabilistic modeling of time-varying uncertainty; (3) adaptive pulse shape learning.

\subsubsection{Physics Constraints Design}
To ensure model outputs comply with physical laws, this paper designs a multi-level physical constraint system. The first level is the momentum conservation constraint, where impulse during collision processes must satisfy the law of momentum conservation. In the x and y directions respectively:

\begin{equation}
  \label{eq:physic_cons1}
  \begin{aligned} 
  &\int_{t_{\text{start}}}^{t_{\text{end}}} F_x(t) \, dt = P_x = m_{\text{eff}} \Delta v_x \\ 
  &\int_{t_{\text{start}}}^{t_{\text{end}}} F_y(t) \, dt = P_y = m_{\text{eff}} \Delta v_y
  \end{aligned}
\end{equation} 
where $m_{\text{eff}} = \frac{m_1 m_2}{m_1 + m_2}$ is the effective mass, and $\Delta v_x$ and $\Delta v_y$ are the relative velocity changes before and after collision. The corresponding impulse conservation loss function is:
\begin{equation}
  \label{eq:physic_cons2}
  \begin{aligned} 
  \mathcal{L}_{\text{impulse}} = \lambda_{\text{imp}} \left[ \left(\int_{t_{\text{start}}}^{t_{\text{end}}} F_x(t) dt - P_x\right)^2 + \left(\int_{t_{\text{start}}}^{t_{\text{end}}} F_y(t) dt - P_y\right)^2 \right]
  \end{aligned}
\end{equation} 

The second level is the energy consistency constraint, where energy constraints are introduced to ensure energy conversion during collision processes complies with physical laws:
\begin{equation}
  \label{eq:physic_cons3}
  \begin{aligned} 
  \mathcal{L}_{\text{energy}} = \lambda_{\text{eng}} \left|\int_{t_{\text{start}}}^{t_{\text{end}}} \mathbf{F}(t) \cdot \mathbf{v}_{\text{rel}}(t) \, dt - E_{\text{dissipated}}\right|^2
  \end{aligned}
\end{equation} 
where the dissipated energy $E_{\text{dissipated}} = \frac{1}{2} m_{\text{eff}} V_{\text{rel}}^2 (1 - e^2)$, $e$ is the coefficient of restitution which is set 0.5 as normal semi plastic deformation, and $V_{\text{rel}}$ is the magnitude of relative velocity before collision.

\subsubsection{Neural Network Architecture and Time-Varying Probabilistic Modeling}
\begin{figure}[h]
  \centering
  \includegraphics[width=14.3cm]{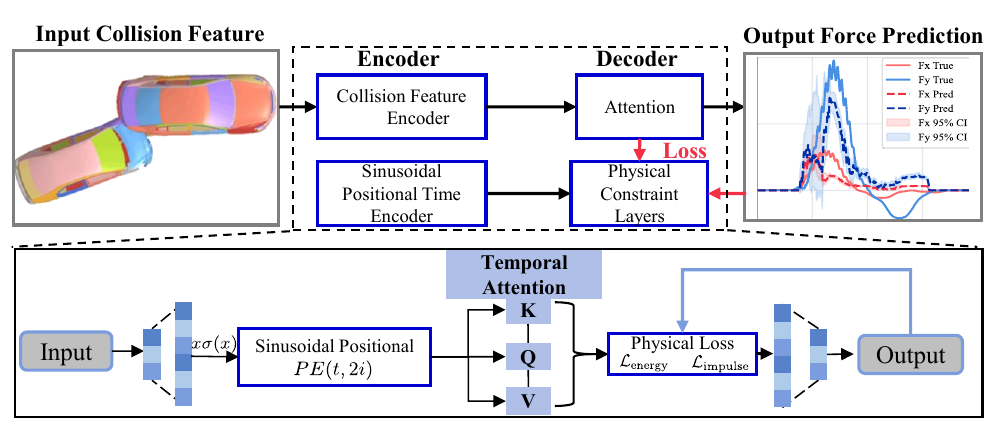}
  \caption{Scheme of PINN-GMM neural network framework}
  \label{fig:nnframework}
\end{figure}

As shown in the Figure \ref{fig:nnframework}, the network adopts an encoder-decoder architecture, primarily comprising four core components with deep integration. The multi-scale feature encoder processes the input collision scenario features $\boldsymbol{\theta}$, employing a multi-branch structure to extract feature representations at different scales: $\mathbf{h}_{\text{feat}} = \text{Swish}(\mathbf{W}_3 \cdot \text{Swish}(\mathbf{W}_2 \cdot \text{Swish}(\mathbf{W}_1 \cdot \boldsymbol{\theta})))$. The sinusoidal positional time encoder employs sinusoidal positional encoding to process temporal information, enhancing the network's understanding of temporal patterns: $\text{PE}(t, 2i) = \sin\left(\frac{t}{10000^{2i/d}}\right)$, $\text{PE}(t, 2i+1) = \cos\left(\frac{t}{10000^{2i/d}}\right)$. This encoding approach enables the network to better capture the periodicity and temporal dependencies of impact forces. The temporal attention mechanism introduces self-attention to enhance the network's focus on critical temporal nodes: $\text{Attention}(Q, K, V) = \text{softmax}\left(\frac{QK^T}{\sqrt{d_k}}\right)V$. The physical constraint integration layer serves as the output layer design, ensuring final outputs satisfy physical laws.

The network employs the Swish activation function: $\text{Swish}(x) = x \cdot \sigma(x) = \frac{x}{1 + e^{-x}}$. The Swish function exhibits excellent numerical stability and gradient propagation characteristics, particularly suitable for training deep physical networks. Compared to traditional ReLU or Tanh functions, Swish demonstrates superior convergence performance in handling physics-constrained optimization. The network output layer is designed to include three main components: instantaneous impact force prediction $\mathbf{F}(t) = [F_x(t), F_y(t)]^T$, adaptive collision duration $\Delta t = \sigma(z_{\Delta t}) \times \Delta t_{\max}$, and dynamic GMM parameters $\boldsymbol{\Phi} = [\boldsymbol{\pi}, \boldsymbol{\mu}, \boldsymbol{\Sigma}]$.

Finally, this paper designs a time-varying GMM where parameters are modeled as continuous functions of time and scenario features:
\begin{equation}
  \label{eq:gmmfunction}
  \begin{aligned} 
  p(\mathbf{F}(t)|\boldsymbol{\theta}) = \sum_{k=1}^K \pi_k(t, \boldsymbol{\theta}) \mathcal{N}(\mathbf{F}(t)|\boldsymbol{\mu}_k(t, \boldsymbol{\theta}), \boldsymbol{\Sigma}_k(t, \boldsymbol{\theta}))
  \end{aligned}
\end{equation}
The time-varying weight function $\pi_k(t, \boldsymbol{\theta}) = \frac{\exp(w_k(t, \boldsymbol{\theta}))}{\sum_{j=1}^K \exp(w_j(t, \boldsymbol{\theta}))}$; the time-varying mean function $\boldsymbol{\mu}_k(t, \boldsymbol{\theta}) = \mathbf{F}_{\text{base}}(t, \boldsymbol{\theta}) + \boldsymbol{\delta}_k(t, \boldsymbol{\theta})$ and the time-varying covariance function $\boldsymbol{\Sigma}_k(t, \boldsymbol{\theta}) = \text{diag}(\sigma_{\min}^2 + \exp(\mathbf{s}_k(t, \boldsymbol{\theta})))$. 
This design enables the model to provide dynamically adjusted uncertainty estimates at different time instants and under different collision scenarios, offering significant advantages over traditional static uncertainty modeling approaches.

\subsection{Model Analysis Based on FEA Dataset}
\begin{figure}[h]
  \centering
  \includegraphics[width=13cm]{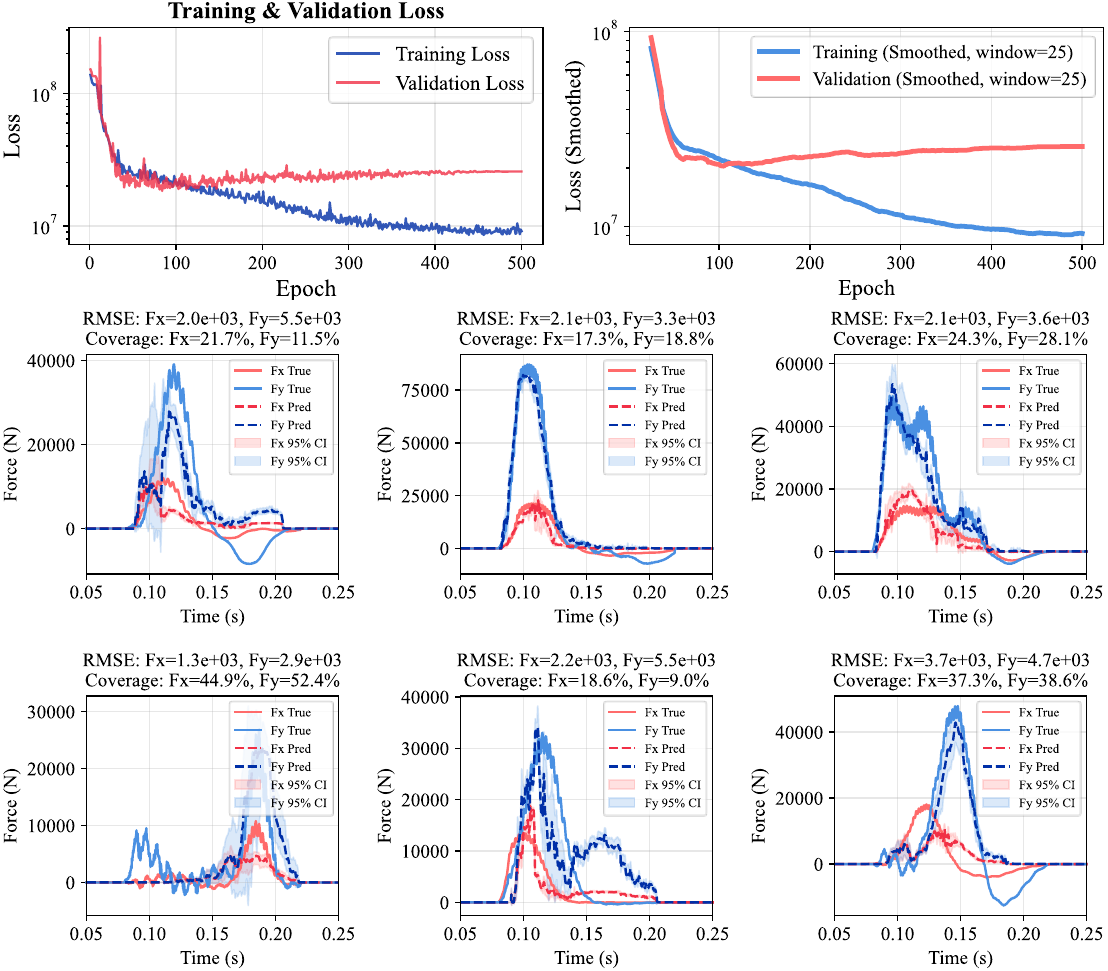}
  \caption{PINN-GMM losses and six samples of force prediction results}
  \label{fig:force_prediction}
\end{figure}

Numerical simulation data from 100 unique collision scenarios are utilized, with the training set and test set split at a 3:1 ratio. The network architecture adopts an 8-layer fully connected network, featuring hidden layer neuron counts of [256, 512, 512, 256, 128, 64, 32, 16]. Training parameters are set as follows: learning rate of 1e-3, batch size of 32, and 500 training epochs. The PINN-GMM-based prediction model is trained on a computing platform equipped with an NVIDIA GeForce RTX 4070 graphics card. As illustrated in the Figure \ref{fig:force_prediction}, the loss function of the trained model converges and stabilizes at approximately 300 epochs, demonstrating favorable convergence characteristics. Both training loss and validation loss decrease synchronously, with no significant overfitting observed.

As shown in Figure \ref{fig:force_prediction}, all cases in the test set exhibit favorable overall predictions and uncertainty distributions, particularly regarding the trend and magnitude of force peaks. Thus, the proposed model can be deemed to achieve satisfactory performance. The RMSE values of $F_x$ and $F_y$ across all cases are 2.1e+03 and 3.9e+03, respectively, indicating that the model attains a relatively high level of prediction accuracy.




\section{Adaptive PINN Vehicle Dynamics Response Modeling Framework}
Traditional PINN approaches face a fundamental dilemma: rigid physical constraints can completely suppress data-driven learning, effectively reducing neural networks to complex implementations of existing physical models. This paper addresses this critical challenge through a Adaptive Physics Integration mechanism that maintains physical consistency while preserving the network's learning capability.

\subsection{PINN Network Architecture Design}
Following the impact force prediction from the first network, this section designs a second Adaptive physics-informed neural network to model vehicle dynamic responses under uncertain impact forces. Unlike traditional methods that apply physical constraints as post-processing corrections, this framework directly embeds adaptive physical constraints into the network architecture, effectively addressing the core issue of constraint-induced learning inhibition.

The Adaptive PINN can be expressed as:
\begin{equation}
\label{eq:Adaptive_pinn}
\hat{\mathbf{x}}(t+1) = \mathcal{NN}_{\text{Adaptive-PINN}}(\mathbf{F}_{\text{collision}}(t), \boldsymbol{\Theta}_{\text{vehicle}}, t; \mathbf{W}_{\text{adaptive}})
\end{equation}
where $\hat{\mathbf{x}}(t+1) = [v_x(t+1), v_y(t+1), \dot{\psi}(t+1), \dot{\phi}(t+1)]^T$ is the predicted vehicle state vector at time $t+1$, $\mathbf{F}_{\text{collision}}(t) = [F_x(t), F_y(t)]^T$ is the impact force vector from the first PINN network, $\boldsymbol{\Theta}_{\text{vehicle}}$ contains vehicle parameters (mass $m$, moment of inertia $I_{zz}, I_{xx,s}, I_{xz}$, geometric parameters $l_f, l_r, t_w$, and tire parameters), and $\mathbf{W}_{\text{adaptive}}$ represents the adaptive neural network parameters, whose constraint sensitivity is dynamically adjusted based on training progress and data quality.

\begin{figure}[h]
  \centering
  \includegraphics[width=14.3cm]{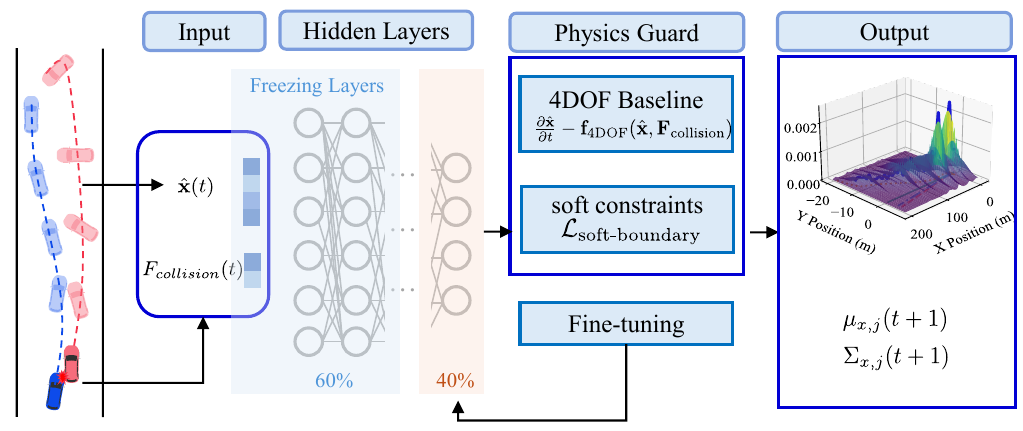}
  \caption{Scheme of A-PINN architecture for post-impact vehicle dynamics prediction}
  \label{fig:P-PINN}
\end{figure}

The vehicle dynamics PINN employs a specialized encoder-decoder architecture optimized for multi-physics constraint integration. As shown in Figure \ref{fig:P-PINN}

First, the input force $\mathbf{F}_{\text{collision}}(t)$ is encoded by the Multi-Scale Feature Encoder module, where the vehicle parameter vector $\boldsymbol{\Theta}_{\text{vehicle}}$ is densely embedded to create a feature representation of the encoded physical properties:
\begin{equation}
  \label{eq:VSD_PINN_swish}
  \mathbf{h}_{\text{vehicle}} = \text{Swish}(\mathbf{W}_3 \cdot \text{Swish}(\mathbf{W}_2 \cdot \text{Swish}(\mathbf{W}_1 \cdot \boldsymbol{\Theta}_{\text{vehicle}})))
\end{equation}
where $\mathbf{W}_1, \mathbf{W}_2, \mathbf{W}_3$ are learnable weight matrices that progressively transform vehicle parameters into high-dimensional feature representations.

The decoder then integrates temporal and vehicle features via a cross-attention mechanism, followed by an output layer of physical constraints:
\begin{equation}
  \label{eq:physics_guard}
  \begin{aligned} 
  \mathbf{h}_{\text{fused}} = \text{CrossAttention}(\mathbf{h}_{\text{combined}}, \mathbf{h}_{\text{vehicle}})
  \end{aligned}
\end{equation}

Traditional PINN methods use fixed constraint weights, which can easily lead to excessive dominance of physical constraints or insufficient data learning. This paper proposes adaptive constraint weights based on training progress:

\begin{equation}
\label{eq:adaptive_weights}
\lambda_{\text{physics}}(t, \text{epoch}) = \lambda_{\min} + (\lambda_{\max} - \lambda_{\min}) \cdot \sigma\left(\frac{\text{epoch} - t_0}{\tau}\right) \cdot \text{DataQuality}(t)
\end{equation}
where $\lambda_{\min} = 0.1$ and $\lambda_{\max} = 10.0$ are the minimum and maximum constraint weights, $\sigma(\cdot)$ is the sigmoid activation function, $t_0 = 300$ indicates the transition round at which the constraint strength begins to decay, $\tau = 100$ controls the smoothness of the transition, and DataQuality$(t)$ measures the local data reliability at time $t$.

Data quality assessment is calculated using nearest neighbor distance:
\begin{equation}
\label{eq:data_quality}
\text{DataQuality}(t) = \exp\left(-\frac{d_{\text{nearest}}(t)}{h_{\text{bandwidth}}}\right)
\end{equation}
where $d_{\text {nearest}}(t)$ is the euclidean distance to the nearest training sample in the temporal feature space at time $t $, and $h_{\text {bandwidth}}= 0.1 $ is the bandwidth parameter that controls the locality of data quality assessment.

The CrossAttention mechanism allows the network to selectively focus on relevant temporal features based on vehicle characteristics, while PhysicsGuard ensures outputs remain within physically meaningful bounds.

\subsection{Adaptive Physics Guard}
To address the fundamental challenge of ensuring physical consistency while enabling the network to learn systematic model discrepancies from limited real data, an Adaptive Physical Guard layer is introduced as the final processing stage. This layer serves as a critical safeguard that prevents physically impossible predictions while preserving the network's ability to capture unmodeled dynamics.

To prevent the vanishing gradients of the model, the physical protection layer is enforced through soft constraints:
\begin{equation}
\label{eq:soft_constraints}
\mathcal{L}_{\text{soft-boundary}} = \lambda_{\text{boundary}} \sum_{i} \text{SoftPlus}\left(\frac{|\hat{x}_i| - b_i}{s_i}\right)^2
\end{equation}
where $\lambda_{\text{boundary}} = 1.0$ is the boundary constraint weight, $\hat{x}_i$ represents the $i$-th predicted state variable, $b_i$ is the physical boundary of the state variable $\hat{x}_i$, $s_i$ is the learnable smooth parameter, and SoftPlus$(\cdot) = \ln(1 + \exp(\cdot))$ ensures the differentiability of the entire constraint region.

To prevent physical constraints from completely suppressing data learning, implement a dynamic balance monitoring system:

\begin{equation}
\label{eq:balance_ratio}
\text{Balance Ratio}(t) = \frac{\mathcal{L}_{\text{data}}(t)}{\mathcal{L}_{\text{data}}(t) + \mathcal{L}_{\text{physics}}(t)}
\end{equation}
where $\mathcal{L}_{\text{data}}(t)$ is the data fit loss at time $t$,$\mathcal{L}_{\text{physics}}(t)$ is the physical constraint loss at time $t$, Balance Ratio quantifies the relative dominance of data versus physics in the current loss landscape, with values close to 0 indicating physics dominance and close to 1 indicating data dominance.

When physical constraints are overly dominant, the system automatically adjusts the constraint strength:
\begin{equation}
\label{eq:auto_adjustment}
\lambda_{\text{physics}} = \lambda_{\text{physics}} \cdot \max(0.1, \text{Balance Ratio})
\end{equation}
where $\lambda_{\text{physics}}$is the current constraint weight, and a factor of 0.1 ensures that a minimum constraint strength is maintained even if data dominance is detected to maintain basic physical consistency.
Physical residual learning can be expressed as:
\begin{equation}
\label{eq:residual_physics}
\mathcal{L}_{\text{physics}} = \lambda_{\text{physics}}(t) \left\|\frac{\partial \hat{\mathbf{x}}}{\partial t} - \mathbf{f}_{\text{4DOF}}(\hat{\mathbf{x}}, \mathbf{F}_{\text{collision}})\right\|^2
\end{equation}
where $\hat{\mathbf{x}}$ is the predicted vehicle state vector, $\mathbf{f}_{\text{4DOF}}(\cdot)$ represents the baseline 4DOF vehicle dynamics function from the equation \ref{eq:4dof}, $\mathbf{F}_{\text{collision}}$ is the crash force input, and $\lambda_{\text{physics}}(t)$ is the time-varying physical constraint weight.

\subsection{Vehicle State GMM Output and Uncertainty Modeling}
Uncertainty in the second PINN model is primarily due to: (1) impact force uncertainty from the first network $\mathbf{F}_{\text{collision}}(t)$;(2) vehicle parameter measurement error $\boldsymbol{\Theta}_{\text{vehicle}}$;(3) unmodeled dynamics of the baseline 4DOF model;(4) model uncertainty due to limited training data; and (5) uncertainty due to adaptive physical constraint weight adjustments.

The network output layer predicts the GMM parameters of the vehicle state vector $\hat{\mathbf{x}}(t+1) = [v_x(t+1), v_y(t+1), \dot{\psi}(t+1), \dot{\phi}(t+1)]^T$, where $v_x(t+1)$ and $v_y(t+1)$ are the longitudinal and lateral speeds, respectively, $\dot{\psi}(t+1)$ is the yaw rate, and $\dot{\phi}(t+1)$ is the roll rate:

\begin{equation}
\label{eq:vehicle_state_gmm}
p(\hat{\mathbf{x}}(t+1)|\mathbf{F}_{\text{collision}}(t), \boldsymbol{\Theta}_{\text{vehicle}}) = \sum_{j=1}^{J} \omega_j(t) \mathcal{N}(\hat{\mathbf{x}}(t+1)|\boldsymbol{\mu}_{x,j}(t+1), \boldsymbol{\Sigma}_{x,j}(t+1))
\end{equation}
where $J$ is the number of mixed components, $\omega_j(t)$is the time-varying weight of the $j$th component, $\boldsymbol{\mu}_{x,j}(t+1) \in \mathbb{R}^4$ is the mean vector of the $j$th component, and $\boldsymbol{\Sigma}_{x,j}(t+1) \in \mathbb{R}^{4 \times 4}$ is the covariance matrix of the $j$th component.

The original network outputs $z_{\omega,j}, \mathbf{z}_{\mu,j}, \mathbf{z}_{\sigma,j}$ are transformed as follows to ensure a valid probability distribution:
\begin{equation}
\label{eq:gmm_parameters_transformation}
\begin{aligned}
\omega_j(t) &= \frac{\exp(z_{\omega,j})}{\sum_{k=1}^{J} \exp(z_{\omega,k})} \\
\boldsymbol{\mu}_{x,j}(t+1) &= \mathbf{f}_{\text{4DOF}}(\hat{\mathbf{x}}(t), \mathbf{F}_{\text{collision}}(t)) + \tanh(\mathbf{z}_{\mu,j}) \odot \boldsymbol{\sigma}_{\text{bound}} \\
\boldsymbol{\Sigma}_{x,j}(t+1) &= \text{diag}(\sigma_{\text{min}}^2 + \exp(\mathbf{z}_{\sigma,j}))
\end{aligned}
\end{equation}
where the first equation ensures that the sum of weights is 1 and non-negative through the softmax function; The second equation centers on the 4DOF physics model prediction and limits the learning bias to $[-1,1]$ via the $\tanh(\cdot)$ function, where $\boldsymbol{\sigma}_{\text{bound}} = [5.0, 3.0, 1.0, 0.8]^T$ is the maximum allowable deviation of $[v_x, v_y, \dot{\psi}, \dot {\phi}]^T$, respectively; the third equation ensures positive definite covariance matrix by diagonalization and exponential function, $\sigma_{\text{min}} = 0.01$ prevents numerical instability.

Training objectives combine data fitting and physical consistency, ensuring that each GMM component satisfies the physical laws:
\begin{equation}
\label{eq:gmm_physics_loss}
\mathcal{L}_{\text{GMM-physics}} = \lambda_{\text{phys-gmm}} \mathcal{L}_{\text{physics-gmm}} + \lambda_{\text{consistency}} \mathcal{L}_{\text{consistency}}
\end{equation}

Physical consistency loss ensures that the mean of each GMM component follows the vehicle dynamics equation:
\begin{equation}
\label{eq:physics_gmm_loss}
\mathcal{L}_{\text{physics-gmm}} = \sum_{j=1}^{J} \omega_j(t) \left\|\frac{\partial \boldsymbol{\mu}_{x,j}}{\partial t} - \mathbf{f}_{\text{4DOF}}(\boldsymbol{\mu}_{x,j}, \mathbf{F}_{\text{collision}}) - \boldsymbol{\delta}_{\text{learned},j}\right\|^2
\end{equation}
where $\frac{\partial \boldsymbol{\mu}_{x,j}}{\partial t}$ is the time derivative of the mean of the $j$ component, $\mathbf{f}_{\text{4DOF}}(\cdot)$ is the 4DOF vehicle dynamics function (from the equation \ref{eq:4dof}), $\boldsymbol{\delta}_{\text{ learned},j}$ is the learnable physics correction for the $j$ component to capture unmodeled dynamics.

Preventing GMM component over-clustering due to diversity consistency loss:
\begin{equation}
\label{eq:consistency_loss}
\mathcal{L}_{\text{consistency}} = \sum_{j=1}^{J-1} \sum_{k=j+1}^{J} \exp\left(-\frac{\|\boldsymbol{\mu}_{x,j} - \boldsymbol{\mu}_{x,k}\|^2}{2\tau^2}\right)
\end{equation}
where $\tau = 2.0$ is the bandwidth parameter that controls the degree of separation between components. Losses increase when two components are too close together,  encouraging diversity.

\subsection{Uncertainty Propagation and Trajectory Distribution Modeling}
Vehicle state uncertainty is propagated to the global trajectory coordinate system by traceless transformation. For each GMM component $j$, generate $2n+1$ sigma points from its mean $\boldsymbol{\mu}_{x,j}(t+1)$and covariance $\boldsymbol{\Sigma}_{x,j}(t+1)$(where $n=4$is the state dimension):
\begin{equation}
\label{eq:sigma_points_gmm}
\begin{aligned}
\boldsymbol{\chi}_{j,0} &= \boldsymbol{\mu}_{x,j}(t+1) \\
\boldsymbol{\chi}_{j,i} &= \boldsymbol{\mu}_{x,j}(t+1) + \left(\sqrt{(n+\lambda)\boldsymbol{\Sigma}_{x,j}(t+1)}\right)_i, \quad i = 1,\ldots,n \\
\boldsymbol{\chi}_{j,i} &= \boldsymbol{\mu}_{x,j}(t+1) - \left(\sqrt{(n+\lambda)\boldsymbol{\Sigma}_{x,j}(t+1)}\right)_{i-n}, \quad i = n+1,\ldots,2n
\end{aligned}
\end{equation}
where $\lambda =\alpha ^2 (n+\kappa) - n$ is the scaling parameter, $\alpha = 0.001$ controls the sigma point distribution, $\kappa = 0$ is the secondary scaling parameter, $\left(\sqrt{(n+\lambda)\boldsymbol{\Sigma}_{x,j}(t+1)}\right)_i$ represents the $i$ column of the square root of the matrix.

Each sigma point is transformed into the global coordinate system through kinematic equations:
\begin{equation}
\label{eq:sigma_transform}
\mathcal{Y}_{j,i} = \begin{bmatrix} X(t) + v_{x,i} \cos \psi(t) \Delta t - v_{y,i} \sin \psi(t) \Delta t \\ Y(t) + v_{x,i} \sin \psi(t) \Delta t + v_{y,i} \cos \psi(t) \Delta t \end{bmatrix}
\end{equation}
where $X(t), Y(t)$ is the current global position, $\psi(t)$ is the current heading angle, $v_{x,i}, v_{y,i}$ is the velocity component of the $i$th sigma point, and $\Delta t$ is the time step.

The parameter distribution of the transformed trajectory is calculated through weighted statistics:
\begin{equation}
\label{eq:trajectory_moments}
\begin{aligned}
\boldsymbol{\mu}_{\text{traj},j}(t+1) &= \sum_{i=0}^{2n} W_i^{(m)} \mathcal{Y}_{j,i} \\
\boldsymbol{\Sigma}_{\text{traj},j}(t+1) &= \sum_{i=0}^{2n} W_i^{(c)} (\mathcal{Y}_{j,i} - \boldsymbol{\mu}_{\text{traj},j}(t+1))(\mathcal{Y}_{j,i} - \boldsymbol{\mu}_{\text{traj},j}(t+1))^T
\end{aligned}
\end{equation}
where the weights are defined as: $W_0^{(m)} = \lambda/(n+\lambda)$, $W_i^{((m)} = 1/(2(n+\lambda)))$ $(i=1,\ldots,2n)$, $W_0^{(c)} = \lambda/(n+\lambda) + (1-\alpha^2+\beta)$, $W_i^{(c)} = 1/(2(n+\lambda))$ $ (i=1,\ldots,2n)$,$\beta = 2$ is the higher-order moment correction parameter.

The final global trajectory GMM distribution combines all transformation components:
\begin{equation}
\label{eq:complete_trajectory_gmm}
p(X(t+1), Y(t+1)|\mathbf{F}_{\text{collision}}(t)) = \sum_{j=1}^{J} \omega_j(t) \mathcal{N}([X(t+1), Y(t+1)]^T; \boldsymbol{\mu}_{\text{traj},j}(t+1), \boldsymbol{\Sigma}_{\text{traj},j}(t+1))
\end{equation}

\subsection{Training Strategy}
The complete training loss integrates multiple physical and statistical objectives to ensure both accuracy and physical consistency:
\begin{equation}
  \label{eq:all_loss}
  \begin{aligned} 
  &\mathcal{L}_{\text{total}} = \mathcal{L}_{\text{data}} + \lambda_{\text{phys}} \mathcal{L}_{\text{physics-GMM}} + \lambda_{\text{traj}} \mathcal{L}_{\text{trajectory}} + \lambda_{\text{smooth}} \mathcal{L}_{\text{smoothness}} + \lambda_{\text{bound}} \mathcal{L}_{\text{soft-boundary}} \\
  &\mathcal{L}_{\text{data}} = ||\hat{\mathbf{x}}(t+1) - \mathbf{x}_{\text{true}}(t+1)||^2\\
  &\mathcal{L}_{\text{smoothness}} = \left|\left|\frac{\partial^2\hat{\mathbf{x}}}{\partial t^2}\right|\right|^2
  \end{aligned}
\end{equation}
where the weight coefficients $\lambda_{\bullet}$ control the relative importance of each loss component during training. $\mathbf{x}_{\text{true}}$ is the true data. $\mathbf{x}_0$ is the known initial vehicle state at collision onset time $t_0$.
The PINN model proposed in this paper adopts a pre-training and fine-tune approach, and the physical prior knowledge or data knowledge based on true values used by all the compared Models are relatively fair.

The initial pre-training phase focuses on embedding basic physical constraints, with training more considering physics loss to enable the network to learn vehicle dynamic physical laws. 100 force data obtained from FEA are input into a 4DOF model for basic simulation, and the output of the 4DOF model is used as the basis for PINN pre-training, where the trajectory sampling frequency is 100Hz and consistent with the data frequency used for fine-tuning with Carsim data.

The fine-tune leverages the physically consistent foundation established during pre-training, achieving optimization performance with less training data through selective layer freezing and mixed loss optimization. In this paper, 5 and 20 real Carsim simulation trajectories which contain 2000 and 8000 samples are used for training during the fine-tune.

To prevent overfitting and preserve learned physics while adapting to limited real data, a layer freezing approach is employed:
\begin{equation}
  \label{eq:layer_freeze}
 \mathbf{W}_{\text{fine-tune}} = \{\mathbf{W}_{\text{frozen}}, \mathbf{W}_{\text{active}}\}
\end{equation}
where $\mathbf{W}_{\text{frozen}}$ represents the frozen parameters from pre-training that encode fundamental physics relationships, and $\mathbf{W}_{\text{active}}$ represents the active parameters that adapt to specific collision scenarios. The number of frozen layers $N_{\text{frozen}}$ is determined as $N_{\text{frozen}} = \lfloor 0.6 \times N_{\text{total}} \rfloor$. Ensuring that 60\% of layers remain frozen to preserve physics knowledge while allowing 40\% to adapt to trajectory-specific patterns. 


\subsection{Baseline Models}
To validate the proposed PINN method, a comprehensive comparison was conducted with two baseline methods. Notably, the comparison between different models is based on the premise that the prior physical knowledge and true data are completely consistent, where 20\% of the data obtained from the dataset was selected as known data.
A conventional numerical integration approach using the same 4DOF vehicle dynamics equations without neural network approximation serves as the 4DOF baseline model:
\begin{equation}
  \label{eq:baseline:4dof}
  \mathbf{x}(t+\Delta t) = \mathbf{x}(t) + \mathbf{f}(\mathbf{x}(t), \mathbf{u}(t), \mathbf{F}_{\text{collision}}(t)) \cdot \Delta t
\end{equation}
where $\mathbf{f}(\cdot)$ represents the system dynamics function derived from the 4DOF equations, and $\mathbf{u}(t)$ represents control inputs (steering, throttle, braking). For basic and easily obtainable vehicle parameters, they are directly exported from the Carsim model, while for tire parameters $K_i, C_i$ that have a significant dynamic impact on the vehicle under extreme conditions, particle swarm optimization (PSO) is used to obtain the best matching parameters from 8000 prior sampling data. 


Another baseline model is a standard feedforward neural network without physical constraints, trained purely based on input-output mapping, removing the Physics Guard layer in PINN and only considering data loss during training, representing a purely empirical method for fitting the data:
\begin{equation}
  \label{eq:baseline:pure_nn}
  \hat{\mathbf{x}}_{\text{data}}(t+1) = \mathcal{NN}_{\text{data}}(\mathbf{F}_{\text{collision}}(t), \boldsymbol{\Theta}_{\text{vehicle}}, t; \mathbf{W}_{\text{data}})
\end{equation}

\subsection{Carsim-FEA Force Dataset Combined Simulation Analysis}

By inputting transient impact forces obtained from FEA software into the Carsim model, the corresponding vehicle trajectory is obtained. First, the vehicle's dynamic characteristics under different forces are analyzed based on the global X-Y trajectory. A clustering method for vehicle dynamics simulation data based on force integrals ($P_x, P_y$) is employed. This method, grounded in physical principles, uses $P_x, P_y$ as the core classification criteria. First, four basic force integral features are extracted for each simulation case: X- and Y-direction force integrals ($I_{fx}, I_{fy}$), total impulse ($I_{total} = \int_0^T |F(t)| dt$), and force statistics (maximum value, mean, and standard deviation). The clustering process utilizes an improved K-means algorithm, adaptively determining the optimal number of clusters using the silhouette coefficient. 20 random initializations and a maximum of 500 iterations are used to ensure result stability. Finally, post-processing merges small clusters ($\leq 5$\%) to ensure the physical plausibility of the clustering results.

\begin{figure}[h]
  \centering
  \includegraphics[width=14cm]{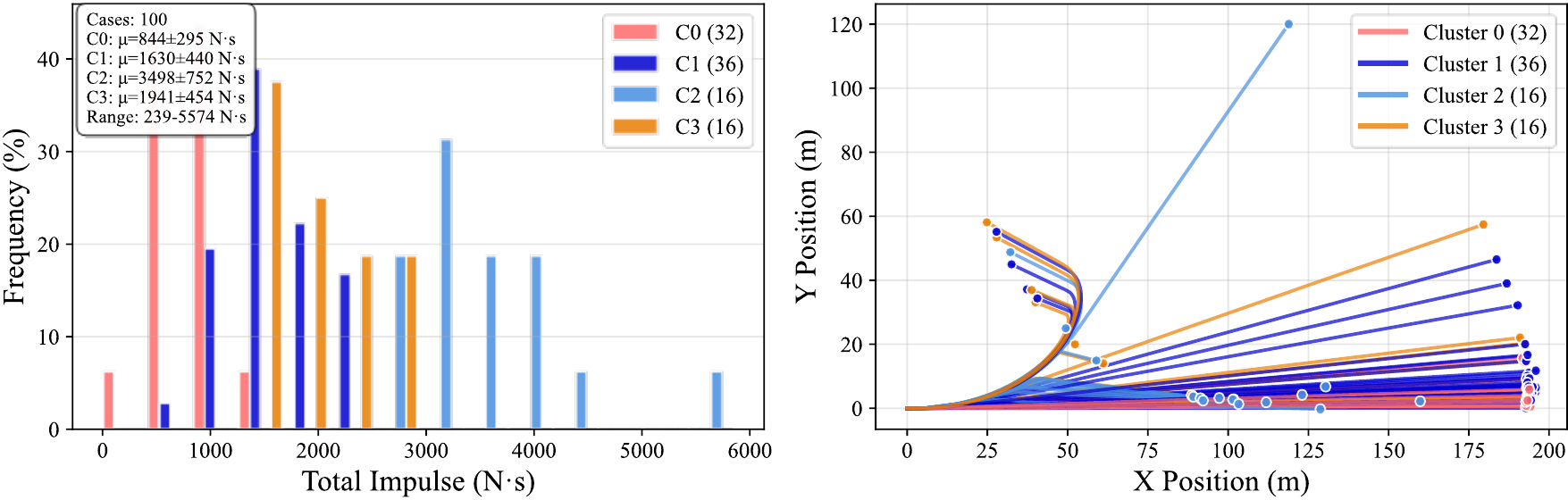}
  \caption{Comparing PINN-based vehicle state prediction with baselines under impact force with uncertainties}
  \label{fig:sim_results_casestudy_all}
\end{figure}

As shown in Figure \ref{fig:sim_results_casestudy_all}, cluster analysis divides the 100 Carsim simulation cases into 4 categories based on impact characteristic patterns, and different clusters exhibit significant differences in X-Y trajectory features. Among them, the vehicles in C0-C4 experience gradually increasing impact, where most trajectories in C0 and C1 show that the vehicles do not spin significantly under impact force, whereas all vehicles in C2 and C3 exhibit clear spinning, leading to loss of control in the latter half of the vehicle trajectories. Since the tire characteristics of the vehicle in these two scenarios are quite different, and tire characteristics are precisely the hard-to-model parts in vehicle dynamics, the following section will present the specific performance of different models in these two types of cases.


Based on the performance comparison analysis in the Table \ref{tab:model_comparison}, different models exhibit distinct characteristics and trade-offs in terms of tracking accuracy, computational efficiency, and physical consistency. The NN-WITHOUT-PINN model ranks first overall with the lowest average error (8.25m) and demonstrates superior performance in most RMSE metrics including X, Y positions and vx velocity, while maintaining the highest computational efficiency (7.45$\pm$1.08 ms/step). However, this purely data-driven model lacks physical constraints, potentially leading to predictions that violate fundamental vehicle dynamics principles and pose safety risks in complex scenarios due to unrealistic outputs. The traditional 4DOF model, despite its fast computation speed (3.21$\pm$2.03 ms/step), suffers from poor adaptability due to fixed parameter settings and exhibits the worst performance across all accuracy metrics, particularly with X-direction position errors reaching 37.75$\pm$29.47m, failing to meet high-precision tracking requirements.

In contrast, the PINN-GMM-20TRAC model demonstrates the best comprehensive performance and practical value. While its average error (10.47m) is slightly higher than the pure neural network approach, it excels in critical parameters for vehicle stability control: vy velocity prediction error of only 1.54$\pm$1.87 km/h and yaw rate prediction error of 2.62$\pm$2.68 deg/s, significantly outperforming other models. More importantly, PINN-GMM-20TRAC incorporates physics-informed neural network constraints to ensure predictions comply with vehicle dynamics principles, maintaining physical reasonableness while achieving high accuracy. The model achieves an optimal balance between computational efficiency and prediction precision, offering better stability and robustness compared to PINN-GMM-5TRAC. This makes it the optimal choice for practical engineering applications that require the integration of accuracy, efficiency, and safety considerations.

\begin{table}[htbp]
\centering
\caption{Model Performance Comparison Analysis}
\label{tab:model_comparison}
\small
\begin{tabular}{|l|c|c|c|c|}
\hline
\multicolumn{5}{|c|}{\textbf{Overall Performance Ranking on Tracking}} \\
\hline
\textbf{Rank} & \textbf{Model} & \textbf{Avg Error (m)} & \textbf{Score} & \textbf{Time (ms)/step} \\
\hline
1 & NN-ONLY & 8.25 & 0.121 & 7.45$\pm$1.08 \\
2 & PINN-20 & 10.47 & 0.096 & 10.78$\pm$1.32 \\
3 & PINN-5 & 14.13 & 0.071 & 10.75$\pm$1.34 \\
4 & 4DOF & 35.20 & 0.028 & 3.21$\pm$2.03 \\
\hline
\hline
\multicolumn{5}{|c|}{\textbf{RMSE Performance (Mean$\pm$Std)}} \\
\hline
\textbf{Variable} & \textbf{4DOF} & \textbf{PINN-5} & \textbf{PINN-20} & \textbf{NN-ONLY} \\
\hline
$X$(m) & 37.75$\pm$29.47 & 10.77$\pm$25.23 & 7.13$\pm$12.20 & 9.55$\pm$9.22 \\
$Y$(m) & 16.40$\pm$13.66 & 13.76$\pm$46.34 & 9.02$\pm$6.08 & 4.88$\pm$5.29 \\
$v_x$(km/h) & 42.96$\pm$57.71 & 7.78$\pm$13.84 & 7.85$\pm$13.83 & 5.94$\pm$6.30 \\
$v_y$(km/h) & 12.59$\pm$8.32 & 1.36$\pm$1.82 & 1.54$\pm$1.87 & 1.58$\pm$1.89 \\
Yaw Rate(deg/s) & 28.93$\pm$99.78 & 15.03$\pm$100.75 & 2.62$\pm$2.68 & 6.26$\pm$6.35 \\
\hline
\end{tabular}
\end{table}

\begin{figure}[H]
  \centering
  \includegraphics[width=14.3cm]{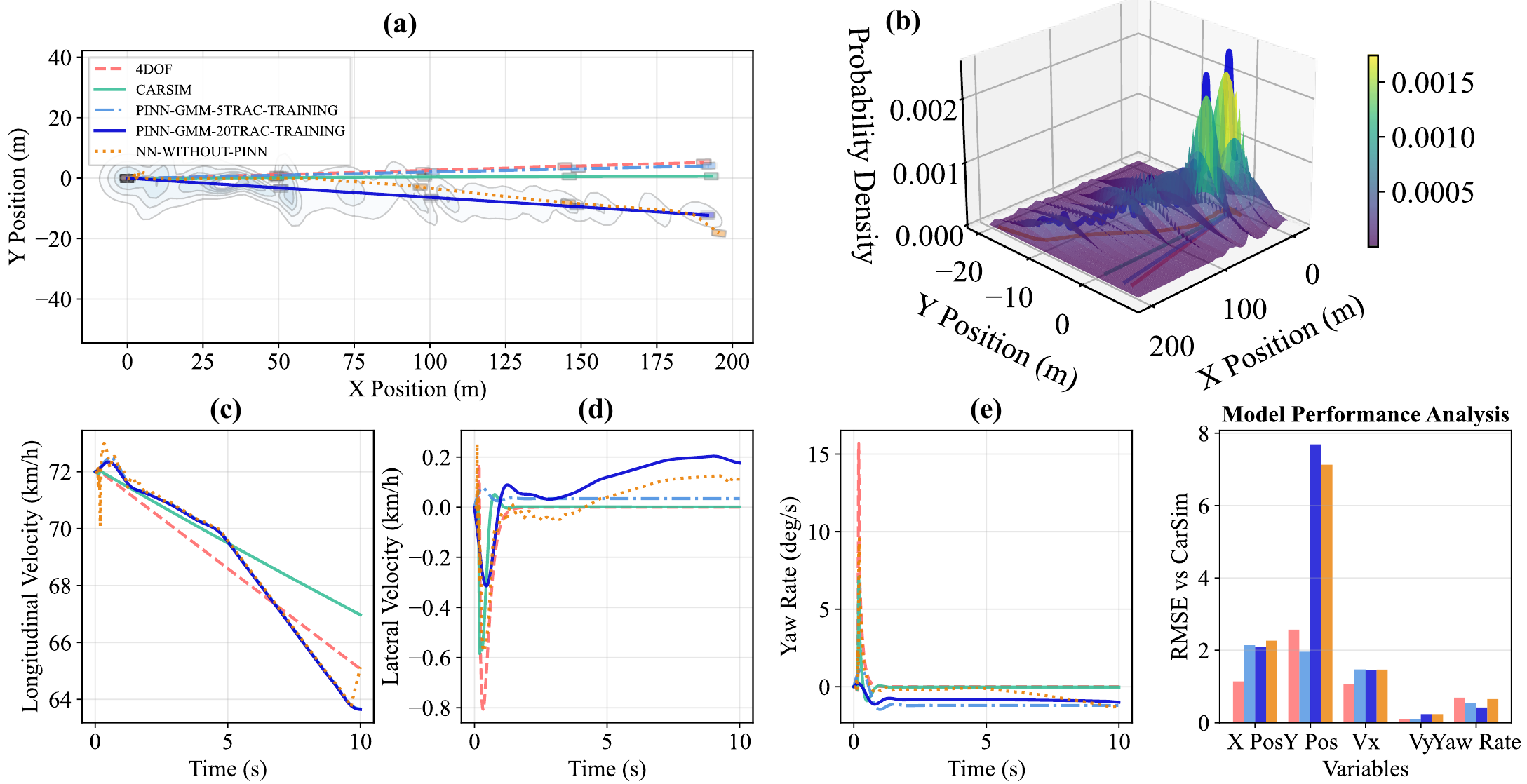}
  \caption{A typical Case for C0}
  \label{fig:sim_results_casestudy2}
\end{figure}

A typical case for C0 is shown in Figure \ref{fig:sim_results_casestudy2}. The 4DOF model in this case exhibits the best accuracy in trajectory tracking, as evidenced by its closest approach to the reference trajectory over the entire path, especially in the challenging curved portion, and achieves the lowest RMSE values on several metrics including position accuracy and yaw rate prediction, since the 4dof model with tires in the linear region theoretically also has excellent performance. However, the PINN-GMM-20 TRAC model has several significant advantages. Most importantly, PINN-based methods provide uncertainty quantification, as shown in (b) uncertainty distribution, where the probability density distribution is centered around the actual trajectory with a peak value of about 0.0015, providing valuable confidence estimates for safety-critical decisions. From this point of view, PINN also has good performance in this case. It is worth noting that training with 5 tracks outperforms training with 20 tracks, which proves that this case is a special case. Although the PINN model showed slightly higher trajectory bias in the latter half of the path, it maintained competitive performance in longitudinal velocity tracking and demonstrated robust handling of complex vehicle dynamics transitions, especially during the initial high yaw rate phase (approximately 15 degrees/sec).

\begin{figure}[H]
  \centering
  \includegraphics[width=14cm]{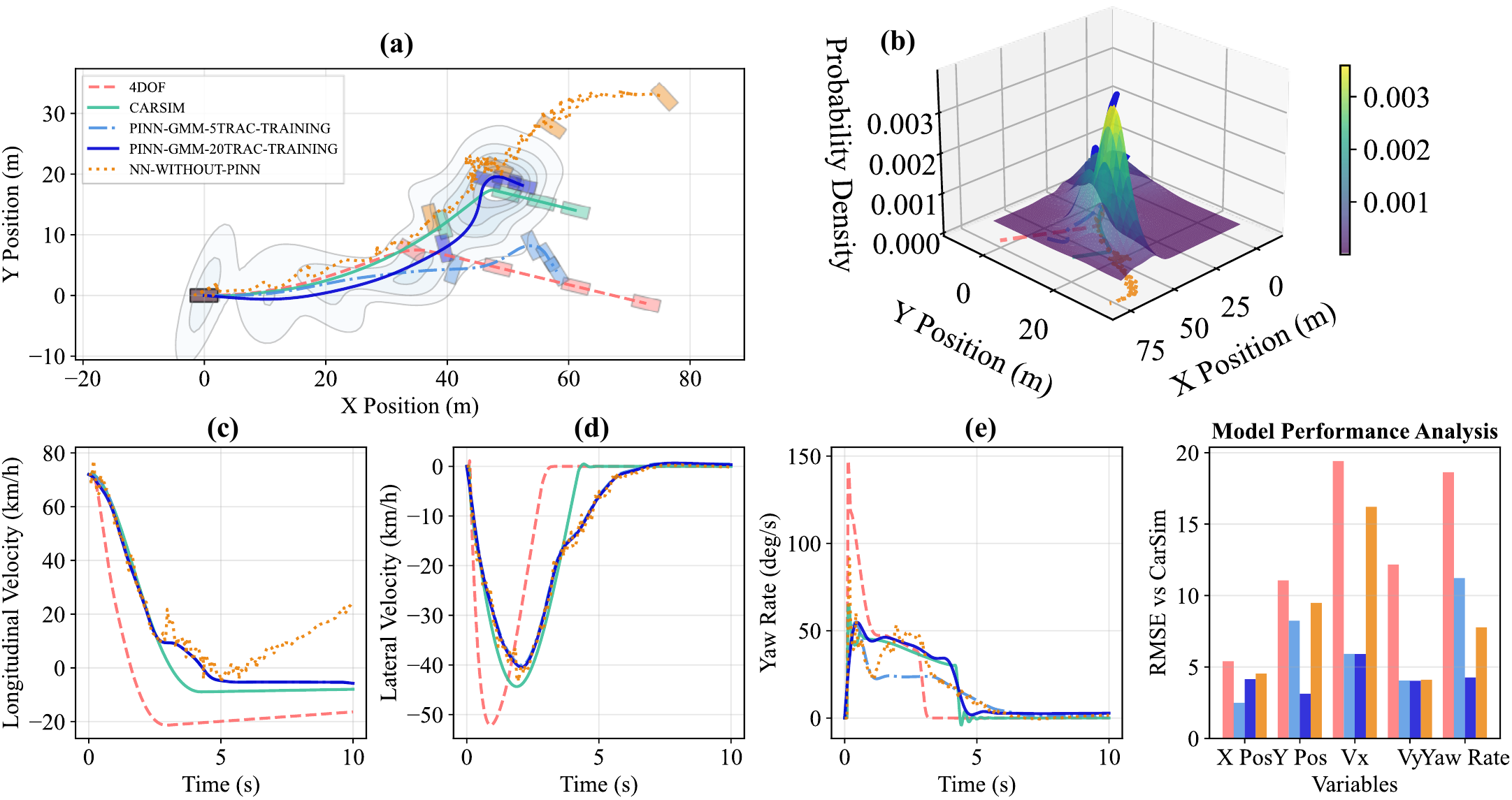}
  \caption{A typical Case for C2}
  \label{fig:sim_results_casestudy1}
\end{figure}

As shown in Figure \ref{fig:sim_results_casestudy1}, the main reason for this is that the tire model of the vehicle has not been estimated well. Throughout the sliding process of the vehicle, the state of the tire keeps changing, and the current state of the vehicle affects the state of the tire. Therefore, the PINN method can better capture the changes in the tire, thus obtaining better vehicle state prediction results, while not deviating from the laws of physics to produce unrealistic scenarios.Pure neural network models completely violate the laws of physics, with trajectories exhibiting situations impossible in real-world scenarios, particularly evident in the increase in longitudinal velocity without force input.

\section{Scaled Vehicle Experiments Results and Analysis}
\subsection{Experiment Setup}
To validate the model's performance on real vehicles, a scaled vehicle experiment is conducted using two chassis platforms that accurately represent vehicle dynamic characteristics. The parameters of the Bullet Vehicle and Target Vehicle are detailed in Table \ref{tab:bullet_para} and Table \ref{tab:Target_para}, respectively. The scale ratio between the experimental vehicles and actual vehicles is approximately 1:4.

\begin{table}[htbp]
\centering
\caption{Bullet Vehicle Parameters}
\label{tab:bullet_para}
\begin{tabular}{|l|l|}
\hline
\textbf{Parameter} & \textbf{Value} \\
\hline
External Dimensions & 2100mm$\times$1120mm$\times$455mm (L$\times$W$\times$H) \\
\hline
Ground Clearance & 120mm \\
\hline
Weight \& Payload & Weight: 290KG \\
\hline
Max Acceleration & 0.8g (No Load) \\
\hline
Rated Power & 12KW \\
\hline
Peak Power & 24KW \\
\hline
Max Speed & 100 km/h \\
\hline
Steering Type & Front-wheel steering\\
\hline
Drive Mode & Independent four-wheel drive\\
\hline
Communication & CAN Bus \\
\hline
\end{tabular}
\end{table}

\begin{table}[htbp]
\centering
\caption{Target Vehicle Parameters}
\label{tab:Target_para}
\begin{tabular}{|l|l|}
\hline
\textbf{Parameter} & \textbf{Value} \\
\hline
External Dimensions & 1900mm$\times$900mm$\times$350mm (L$\times$W$\times$H) \\
\hline
Weight \& Payload & Weight: 100KG\\
\hline
Max Speed & 30 km/h \\
\hline
Steering Type & Front-wheel steering\\
\hline
Drive Mode & rear-wheel drive\\
\hline
Communication & Serial RS232 \\
\hline
\end{tabular}
\end{table}

%
%

\begin{figure}[h]
  \centering
  \includegraphics[width=14cm]{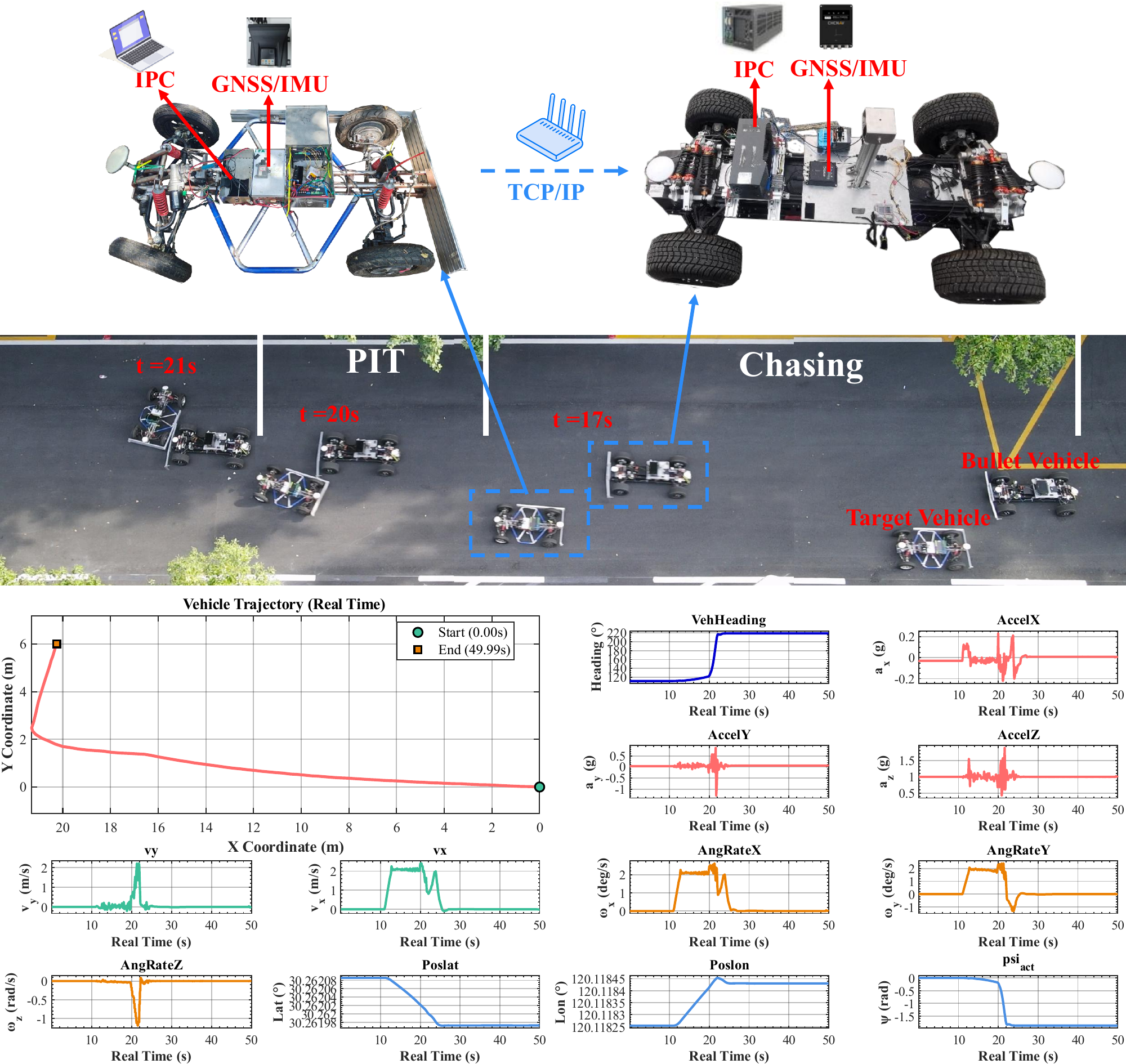}
  \caption{Scaled Vehicles and  Experiment Scenario}
  \label{fig:scaled_vehicle_experiment}
\end{figure}

As shown in Figure \ref{fig:scaled_vehicle_experiment}, PIT experiments are conducted to validate the feasibility and accuracy of the proposed modeling method. Both vehicles used in the experiment are equipped with steer-by-wire and drive-by-wire capabilities, and each is integrated with an Industrial Personal Computer (IPC) for executing planning and control algorithms. A GNSS-IMU system and a camera are installed for data acquisition. Communication between the two vehicles is achieved via TCP/IP protocol within a local area network for real-time data transmission.

The bullet vehicle operates an autonomous driving algorithm designed to execute the PIT maneuver. Based on the real-time position of the target vehicle, it performs prediction, planning, and control to achieve the desired PIT state before initiating the PIT maneuver. Control commands are directly sent via the CAN bus to assign torque signals to each of the four wheels.

The entire PIT process can be divided into two phases. The first phase is the chasing process, where the primary control objective of the bullet vehicle is to achieve a parallel body motion and matching velocity with the target vehicle, while maintaining a lateral distance that ensures the required collision angle and speed for PIT execution. The second phase involves performing the PIT maneuver, whose main goal is to induce a significant yaw change in the target vehicle, thereby disabling its ability to continue moving forward. The model proposed in this paper primarily predicts the vehicle's posture changes during post-impact under the assumption that the steering wheel and throttle are released, and that no electronic stability control (ESC) or similar systems are active. 

Similar to simulation, first use the 8000 sampling data obtained from simulating a 4DOF model containing force inputs as pre-training data to allow the network to fully learn the physical laws, then use the 100 sampling data of scaled experiment for Fine-tuning the model.




The test results are presented in Table \ref{tab:vehicle_model_comparison}. The proposed Physics-Informed Neural Network demonstrates superior trajectory tracking performance, achieving a 63.6\% reduction in average trajectory error (0.73$\pm$0.58 m vs. 2.01$\pm$1.77 m) compared to the conventional 4DOF model, indicating its enhanced capability in capturing complex vehicle dynamics. Notably, the proposed model excels in longitudinal positioning ($R^2$ = 0.984) and velocity prediction ($R^2$= 0.708), demonstrating successful integration of physical constraints with data-driven learning. However, this improvement in accuracy comes with a modest computational overhead, as the proposed method requires slightly longer inference times (13.76 ms vs. 4.42 ms) than the 4DOF model. The inference time remains stable across various operational conditions, indicating good computational stability.

This trade-off between accuracy and computational cost is reasonable, as more complex models inherently require greater computational resources. The increased inference time of the proposed model is primarily attributed to its neural network architecture, which is predictable and expected. Meanwhile, the model's prediction accuracy can be further improved with additional training data, as validated in previous Carsim simulation experiments. Although the proposed approach maintains high accuracy in short-to-medium term predictions, its physical constraints may become insufficient for extended prediction horizons, particularly when limited data availability hinders neural network convergence.

Therefore, the proposed model is particularly suitable for applications requiring high-precision short-term predictions (such as the PIT post-collision trajectory optimization discussed in this paper), while the traditional 4DOF model retains advantages in real-time control systems where computational efficiency is critical. However, when control systems require calibration through data-driven approaches, the proposed framework offers a valuable solution for model improvement through continuous data integration.

\begin{table}[htbp]
\centering
\caption{Vehicle Dynamics Model Performance Analysis of Scaled Vehicle Experiments}
\label{tab:vehicle_model_comparison}
\small
\begin{tabular}{|l|c|c|}
\hline
\multicolumn{3}{|c|}{\textbf{Overall Performance Ranking on Trajectory Tracking}} \\
\hline
\textbf{Model} & \textbf{Avg Error (m)} & \textbf{Time (ms)/step} \\
\hline
PINN Model & 0.73$\pm$0.58 & 13.76$\pm$1.89 \\
4DOF Model & 2.01$\pm$1.77 & 4.42$\pm$3.51 \\
\hline
\hline
\multicolumn{3}{|c|}{\textbf{RMSE Performance (Mean$\pm$Std)}} \\
\hline
\textbf{Variable} & \textbf{4DOF Model} & \textbf{PINN Model} \\
\hline
$X$ Position (m) & 3.55$\pm$1.91 & 0.82$\pm$0.51 \\
$Y$ Position (m) & 1.87$\pm$1.15 & 1.16$\pm$1.28 \\
Longitudinal Velocity (m/s) & 0.81$\pm$0.34 & 0.33$\pm$0.21 \\
Lateral Velocity (m/s) & 0.21$\pm$0.08 & 0.20$\pm$0.08 \\
Yaw Rate (rad/s) & 0.13$\pm$0.05 & 0.14$\pm$0.06 \\
\hline
\end{tabular}
\end{table}

\begin{figure}[H]
  \centering
  \includegraphics[width=14cm]{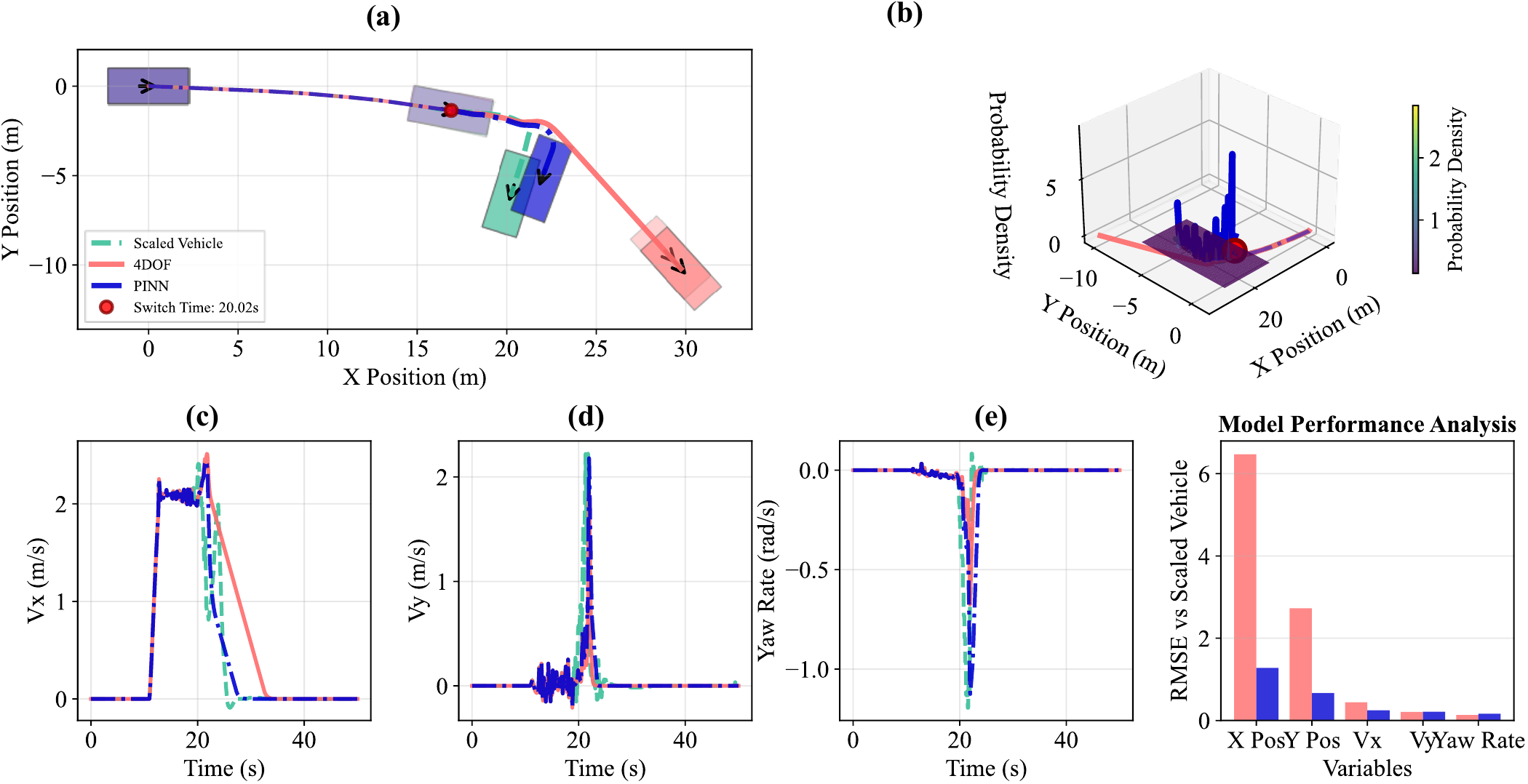}
  \caption{Case 3 Results in Scaled Vehicle Experiments}
  \label{fig:scaled_vehicle_experiment_case3}
\end{figure}

%
As shown in Figure \ref{fig:scaled_vehicle_experiment_case3}, case 3 presents even with this earlier collision time, the proposed model maintains excellent trajectory tracking performance prior to collision. The 4DOF model exhibits more severe deviations in this case, particularly showing substantial errors in Y position predictions. The velocity time series observation indicates that the proposed model more accurately captures the vehicle's dynamic characteristics during steering maneuvers, while the 4DOF model demonstrates obvious deficiencies in lateral dynamics modeling.

\begin{figure}[H]
  \centering
  \includegraphics[width=14cm]{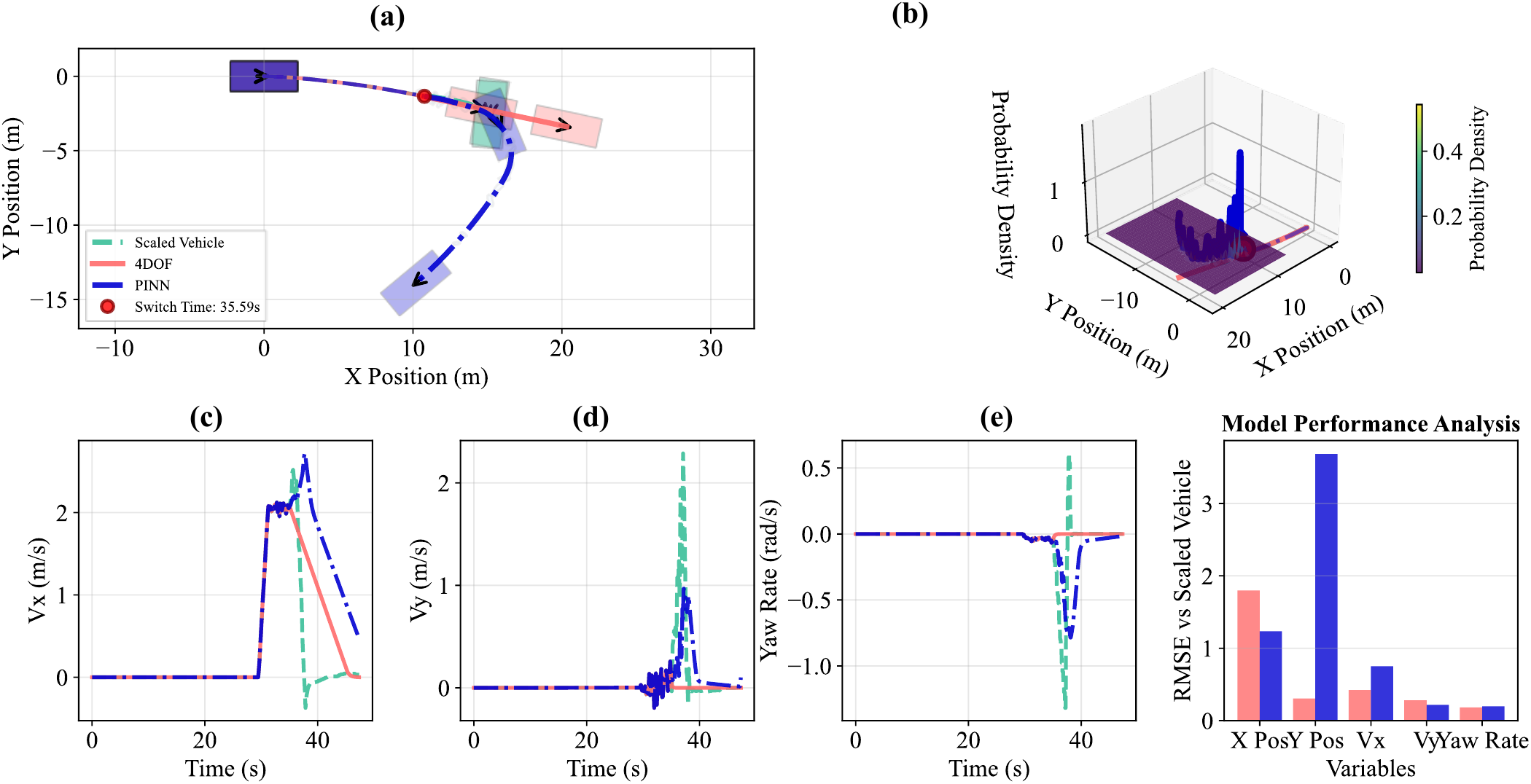}
  \caption{Case 1 Results in Scaled Vehicle Experiments}
  \label{fig:scaled_vehicle_experiment_case1}
\end{figure}

In case 1, as shown in Figure \ref{fig:scaled_vehicle_experiment_case1}, the proposed model demonstrates exceptional trajectory tracking capability, maintaining near-perfect alignment with the reference trajectory until the collision time point at 35.59 seconds, while the 4DOF model shows significant deviation from the beginning. The time series analysis reveals that the proposed model provides more accurate predictions of both $v_x$ and $v_y$. The RMSE comparison demonstrates the proposed model's overwhelming advantage in position prediction, with significantly lower errors than the 4DOF model. Probability density plots further confirm that the proposed model's predictions are more concentrated around the actual trajectory.

Both cases confirm the advantages of the proposed model's physical constraint integration, enabling better consistency with actual vehicle behavior in complex maneuvering scenarios. However, the collision phenomenon observed in the proposed model requires attention, potentially related to numerical stability under extreme conditions and the balance of physical constraint weights. Although the 4DOF model is less accurate than the proposed model, its predictive behavior is more stable without significant numerical divergence. This comparison indicates that model selection should be based on specific accuracy requirements and robustness needs in practical applications. The proposed model shows clear advantages for applications requiring high-precision short-term predictions, while the traditional 4DOF model may be more reliable for control systems requiring long-term stable predictions. Particularly in safety-critical applications, numerical stability and the ability to avoid sudden divergence are crucial.

\section{Conclusion}
This paper proposes two PINN frameworks for vehicle collision dynamics prediction in PIT maneuvers. The first network combines Gaussian Mixture Models with physics constraints to model impact forces, achieving relative errors below 15.0\% on FEA datasets. The second Adaptive PINN predicts post-collision vehicle dynamics with dynamic constraint weighting, reducing average trajectory prediction error by 63.6\% compared to traditional 4DOF model while maintaining millisecond-level computational efficiency.

The framework is validated through computer simulations and scaled vehicle experiments, demonstrating improved accuracy across different test conditions. The method provides uncertainty estimates that are valuable for safety applications, and successfully combines physics knowledge with data-driven learning to work effectively even with limited training data.
Future work should extend the framework to broader collision scenarios, incorporate advanced tire models, and develop automated hyperparameter tuning. This approach bridges theoretical vehicle dynamics with practical collision scenarios, positioning it as a valuable tool for next-generation active safety and post-impact control systems.

Future work should expand the framework to broader collision scenarios and develop automatic parameter tuning methods, while also applying the corresponding models to design autonomous driving algorithms.

\section*{Funding}

This work was supported by National Natural Science Foundation of China (Grant No. 52372421).

\section*{Disclosure statement}
The authors report there are no competing interests to declare.



\section{References}

\bibliography{collisionmodel}

\end{document}